\documentclass[showpacs,preprintnumbers,amsmath,amssymb,floatfix,nofootinbib,aps]{revtex4}

\usepackage[dvips]{graphicx,graphics}
\usepackage{dcolumn}% Align table columns on decimal point
\usepackage{bm}
\usepackage{epsfig}

\newcommand{\MeV}{\;\text{MeV}}
\newcommand{\dirac}{\partial\llap{$\diagup$\kern-2pt}}

\begin{document}

\title{
Roles of axial anomaly on neutral quark matter with
color superconducting phase}

\author{Zhao~Zhang}\email{zhaozhang@pku.org.cn}
\affiliation{ School of Mathematics and Physics, North China
Electric Power University, Beijing 102206, China}

\author{Teiji~Kunihiro}\email{kunihiro@ruby.scphys.kyoto-u.ac.jp}
\affiliation{Department of Physics, Kyoto University,
Kyoto 606-8502, Japan}


\begin{abstract}
We investigate effects of the axial anomaly term with a chiral-diquark
coupling on the phase diagram within a two-plus-one-flavor
Nambu-Jona-Lasinio (NJL) model under the charge-neutrality and
$\beta$-equilibrium constraints. We find that when such constraints
are imposed, the new anomaly term plays a quite similar role
as the vector interaction does on the phase diagram, which the
present authors clarified in a previous work. Thus, there appear several
types of phase structures with multiple critical points at
low temperature $T$,  although the phase diagrams with
intermediate-$T$ critical point(s) are never realized without these constraints
even within the same model Lagrangian.  This drastic change is
attributed to an enhanced interplay between the chiral and diquark
condensates due to the anomaly term at finite temperature;
the u-d diquark coupling is strengthened by the relatively large chiral
condensate of the strange quark through the  anomaly term, which in turn
definitely leads to the abnormal behavior of the diquark condensate at
finite $T$, inherent to the asymmetric quark matter. We note that
the critical point from which the crossover region extends to zero temperature
 appears only when the strength of the vector interaction is larger
than a critical value. We also show that the chromomagnetic instability
of the neutral asymmetric homogenous two-flavor color superconducting(2CSC) phase
is suppressed and can be even completely cured by the enhanced
diquark coupling due to the anomaly term and/or by the vector interaction.
\end{abstract}

\maketitle

%%%%%%%%%%   INTRODUCTION   %%%%%%%%%%

\section{INTRODUCTION}

It is generally believed that the strongly interacting
matter exhibits a rich phase structure in extreme environment
such as at high temperature and high baryon chemical potential.
Experimentally, RHIC (Relativistic Heavy Ion Collider) and LHC
(Large Hadron Collider) may provide more information on
this topics. Theoretically, some results have been already obtained
on a sound basis: First, the lattice simulations of
quantum chromodynamics (QCD) indicate that, for physical quark
masses, the transition from the hadronic phase to the quark gluon
plasma (QGP) is a smooth crossover at finite temperature and
vanishing baryon chemical potential ~\cite{Cheng:2009be,Borsanyi:2010bp},
whereas in the low temperature and very high density region, the techniques
of perturbation QCD can be used and the color flavor locking
(CFL)~\cite{Alford:1998mk} phase is proved to be the ground state
of QCD ~\cite{Son:1998uk,Schafer:1999jg,Shovkovy:1999mr,Schafer:1999fe}.

However, the above methods based on the first principle fail at the
low temperature and moderate density region, due to the sign problem
or the non-perturbative effect. Phenomenologically, such a region
in the $T$-$\mu$ plane is more relevant to reality and hence interesting
since it is directly related to the physics of compact stars. On that
account, chiral  models of QCD such
as the NJL model~\cite{Nambu:1961tp,Vogl:1991qt,Klevansky:1992,Hatsuda:1994pi}
that embody the basic low-energy characteristics
of QCD such as symmetry properties have been extensively used to
explore the $T$-$\mu$ phase diagram of strongly interacting matter.
In particular, such model calculations suggest that CSC phase
may occur at low temperature and large chemical potential (for reviews,
see~\cite{Rajagopal:2000wf, Rischke:2003mt,Buballa:2003qv, Alford:2007xm}).
In addition, a popular result from the model studies is that the chiral phase
transition always keeps first order at the low-temperature region
~\cite{Asakawa:1989bq,Barducci:1989,Kunihiro:1991hp,Berges:1998rc, Ruester:2005jc,Abuki:2005ms}.
Combined with the crossover transition confirmed by lattice QCD at zero baryon chemical
potential, usually, a schematic $T$-$\mu$ phase diagram with one
chiral critical point(CP) is widely adopted in the
literature~\cite{Stephanov:2007fk}. Such a CP may be located at relatively
high temperature and low baryon chemical potential, which has
attracted considerable attention as it is potentially detectable
in heavy-ion experiments~\cite{Stephanov:1998dy,Minami:2009hn}.

Generally, there is no reason to rule out the possibility that the QCD
phase diagram may contain more than one chiral CP, especially
when the chiral and diquark condensates are considered simultaneously;
some rich structures with multiple CP's may be expected for the
phase diagram owing to somehow enhanced interplay between the two
types of condensates\footnote{Note that multiple critical points
had also been found in two-flavor models of QCD without considering
diquark paring\cite{Bowman:2008kc,Ferroni:2010ct}.}.
It has been shown on the basis of the NJL model
that it is indeed the case ~\cite{Kitazawa:2002bc,Addenda,Zhang:2008wx,Zhang:2009mk};
the QCD phase diagram can admit multiple CP's when
the repulsive vector interaction ~\cite{Kitazawa:2002bc,Addenda} or
the charge-neutrality and $\beta$-equilibrium~\cite{Zhang:2008wx}
or both of these two ingredients~\cite{Zhang:2009mk}
are included
\footnote{
Note that the renormalization group theory for deducing low-energy effective
vertex favors the presence of the
vector-type interaction~\cite{ref:EHS,ref:SW-reno}.
}.
This is because the two ingredients act so as to enhance the competition
between the chiral and diquark condensates and thus the would-be first-order
boundary line in the low-temperature region of the $T$-$\mu$ plane can be
turned into a smooth crossover or multiply-cut crossover lines
with new CP('s). Indeed it has been found~\cite{Zhang:2009mk} that
the number of the chiral CP's may vary from zero to four with the
joint effect of these two ingredients. Moreover, the present authors
have shown~\cite{Zhang:2009mk} that the vector interaction can effectively
suppress the chromomagnetic instability \cite{Huang:2004bg}
in the asymmetric homogeneous CSC phase.

It is noteworthy that a direct coupling term between the chiral and
diquark condensates can be supplied by the axial
 anomaly~\cite{Alford:1998mk,Rapp:1999qa,Steiner:2005jm},
which thus might lead to a new CP in the low-temperature
region, as first conjectured in~\cite{Hatsuda:2006ps} on the basis
of an analysis using the Ginzburg-Landau (GL) theory in the chiral
limit; see also subsequent detailed analyses~\cite{Yamamoto:2007ah, Baym:2008me}
though still in the chiral limit. It is,  however, to be noted that the GL theory
assumes that both the diquark  and chiral condensates are sufficiently
small around the phase boundaries provided that the phase transitions are
of second-order. In addition, the GL theory itself can not determine the coefficients
in the action,  and some microscopic model or theory is necessary for such a
determination.

Recently,  a microscopic calculation has been done~\cite{Abuki:2010jq} with
use of a three-flavor NJL model for massive quarks  incorporating the
axial anomaly term with a form of a six-quark
 interaction~\cite{Alford:1998mk,Rapp:1999qa,Steiner:2005jm,Yamamoto:2007ah}
, the coupling constant of which is denoted by $K'$:
It was claimed in ~\cite{Abuki:2010jq} that the low-temperature CP can
exist owing to the axial anomaly for an appropriate range
of the model parameters even off the chiral limit but still with a flavor symmetry
as in the GL approach in the chiral limit. It should be noticed, however, that
the SU(3)-flavor symmetry may lead to a special type of CSC phase, i.e., the CFL
phase, as is taken for granted
in \cite{Hatsuda:2006ps,Yamamoto:2007ah, Baym:2008me,Abuki:2010jq},
 which automatically satisfies the charge-neutrality and
$\beta$-equilibrium constraints.

Then one may suspect that the possible emergence of the new CP might be
an artifact of such an ideal situation with the three-flavor symmetry.
Nevertheless, it is a very interesting possibility that the
axial anomaly-induced interplay between the chiral and diquark
condensates would lead to a new CP in the low-temperature region.
Thus it is worth exploring to see whether a new low-temperature CP
is induced by the axial anomaly in a dynamical model of QCD by
considering the realistic situation with the broken flavor symmetry by the
hierarchical current quark masses.We note that once the quark mass difference
of different flavors is taken into account, it becomes a complicated dynamical
problem to make the charge-neutrality and $\beta$-equilibrium constraints
satisfied.

More recently, such a realistic calculation in the framework of a
two-plus-one-flavor NJL model has been done by Basler
et al.~\cite{Basler:2010xy}; they have shown that such a new low-temperature CP
is not found in such a model even with the axial anomaly term,
because an unusual interplay between the chiral
and diquark condensates induced by the anomaly term   actually
leads favorably to the 2CSC phase~\cite{Alford:1997zt,Rapp:1997zu} rather than
the CFL phase near the chiral phase boundary, even in the case
with the equal quark mass limit ~\cite{Basler:2010xy}.

It is worth emphasizing here that
the constraints by the charge-neutrality and $\beta$-equilibrium are not
taken into consideration in \cite{Abuki:2010jq} nor \cite{Basler:2010xy}
 in contrast to \cite{Zhang:2008wx,Zhang:2009mk} where various types of
multiple-CP structures are found in the phase diagram.
Thus the following two questions arise naturally:
Will the results found in \cite{Basler:2010xy}
be altered or not when the charge-neutrality and $\beta$-equilibrium
constraints and/or the vector interaction are taken into account?
Or will the phase structure with multiple
CP's found in ~\cite{Zhang:2009mk} rather persist when taking
into account the coupling between the chiral and diquark condensates
induced by such a six-quark interaction?
The main purpose of this paper is to answer these questions
by incorporating the  anomaly term that breaks the $U_A(1)$ symmetry
as well as the vector interaction under
the constraints of the charge-neutrality and $\beta$-equilibrium
in the two-plus-one-flavor NJL model.
The present work may be regarded as either an extension of the
paper ~\cite{Basler:2010xy} by incorporating the
charge-neutrality, $\beta$-equilibrium and the vector interaction, or
an extension of the paper~\cite{Zhang:2009mk} by including the
$K'$-interaction.

The main conclusion we reach is that the key results on the phase
structure obtained in Ref.~\cite{Zhang:2008wx,Zhang:2009mk} persist even
when the attractive $K'$-interaction is incorporated. That is,
there appear new CP('s) at the intermediate temperature owing to
charge-neutrality constraint and then the transition in the low
temperature region extending to zero $T$ becomes a crossover  when the
strength of the vector interaction becomes larger than a critical value:
Thus the number of the CP's can be even more than two, depending on
the values of some related coupling constants. Strikingly enough, we
find that the interplay between the chiral and the diquark
condensates induced by the anomaly term even acts toward realizing
the multi-CP structure of the phase diagram under the neutrality
and $\beta$-equilibrium constraints even without the help of the vector
interaction. Accordingly, the results in Ref.~\cite{Basler:2010xy} are
modified by considering these constraints. Even though
the chiral boundary in the low-$T$ region extending to zero $T$
 also remains first order in our case in the absence of the
vector interaction, which shrinks and vanishes eventually as $K'$
becomes large and exceeds a critical value.

We shall also examine the chromomagnetic (in)stability under the influence
of the axial anomaly, as was done in \cite{Zhang:2009mk}.
It is well known that the asymmetric homogenous 2CSC phase
suffers from the chromomagnetic instability. At zero
temperature, the calculation based on the hard-dense-loop (HDL)
method~\cite{Huang:2004bg} suggests that the  Meissner
mass squared of the 8th gluon becomes negative for $\frac{\delta\mu}{\Delta}>1$
while the 4th-7th gluons acquire negative Meissner masses
squared for $\frac{\delta\mu}{\Delta}>1/\sqrt{2}$; here
$\delta \mu$ and $\Delta$ denote the difference of the chemical
potentials of u and d quarks and the gap, respectively. Note that
$\delta\mu$ is just equal to a half of the electron chemical
potential $\mu_e$ when the vector interaction is absent, and this quantity
is to be replaced by an effective chemical potential $\delta\tilde{\mu}$ (see below)
when the vector interaction is present, as shown in \cite{Zhang:2009mk}.
The instability of the asymmetric homogenous CSC phase should imply
the existence of a yet unknown but stable phase in this region of
the $T$-$\mu$ plane. Candidates of such a stable phase include
the Larkin-Ovchinnikov-Fulde-Ferrel (LOFF) phase \cite{Giannakis:2004pf}
and gluonic phase \cite{Gorbar:2005rx}. Besides developing the possible new
phases, the instability problem may also be totally or partially gotten
rid of  by some other mechanisms. For instance, the instability problem
becomes less severe simply at finite temperature because the smeared Fermi
surface relaxes the mismatch of the Fermi spheres of the asymmetric quark matter
\cite{Kiriyama:2006jp,He:2007cn,Fukushima:2005cm}. Furthermore, it is known that
the larger the quark mass and the stronger the diquark coupling, more suppressed
the instability even at zero temperature\cite{Kitazawa:2006zp}. Recently, the present authors\cite{Zhang:2009mk} have shown that the repulsive vector interaction can
also resolve the instability problem totally or partially.
The stability by the vector interaction is realized due to the following
two ingredients:
(1)~the density difference between the u and d quarks reduces the mismatch in the
effective chemical potentials;
(2)~the nonzero vector interaction suppresses the formation of high density and
hence larger quark masses than those obtained without the interaction are realized.
We shall show that the new anomaly term play a quite similar role as the vector
interaction and the interplay between the chiral and diquark condensates induced
by the anomaly term acts toward suppressing the unstable region of the homogeneous
2CSC phase in the $T$-$\mu$ plane; the neutral 2CSC phase can become even free from
the chromomagnetic instability if $K'$ is larger than a critical value $K'_c$, which
can be reduced significantly when the vector interaction is incorporated.

This paper is organized as follows. In Sec.II, the two-plus-one-flavor
NJL model with the extended flavor-mixing six-quark interaction
is introduced under the constraints of the charge-neutrality
and $\beta$-equilibrium. The phase diagram of the neutral strongly
interacting matter with the influence of the axial anomaly is presented
in Sec.III. Sec.IV focuses on the role  of the axial anomaly on the
chromomagnetic (in)stability. The conclusion and outlook are given
in Sec.V.

%%%%%%%%%%   MODEL AND FORMALISM   %%%%%%%%%%
\section{NJL Model With Axial Anomaly and Vector Interaction}
\label{sec: Two-plus-one }

\subsection{The model Lagrangian}

We start from the following two-plus-one-flavor NJL model
with the vector interaction~\cite{Klimt:1989pm, Zhang:2009mk}
and two types of six-quark anomaly terms,
\begin{equation}
\mathcal{L}=\bar \psi \, ( i \dirac - \hat{m} \, )
\psi+\mathcal{L}_{\chi}^{(4)}+\mathcal{L}_{d}^{(4)}+\mathcal{L}_{\chi}^{(6)}
+\mathcal{L}_{\chi{d}}^{(6)},\label{eqn:Lagrangian}
\end{equation}
where
$\hat{m} =\text{diag}_{f}(m_u,m_d,m_s)$ denotes the current-quark mass matrix
and
\begin{eqnarray}
\mathcal{L}_{\chi}^{(4)}=G_S \sum_{i=0}^8 \left[ \left( \bar \psi
\lambda_i^f \psi \right)^2 + \left( \bar \psi i \gamma_5
\lambda_i^f \psi \right)^2 \right]-G_V\sum_{i=0}^8 \left[ \left(
\bar \psi \gamma^\mu \lambda_i^f \psi \right)^2 + \left( \bar \psi
\gamma^\mu \gamma_5 \lambda_i^f \psi \right)^2 \right],\label{eqn:Lagrangian1}\\
\mathcal{L}_{d}^{(4)}=G_D\sum_{i,j=1}^3 \left[(\bar{\psi} i
\gamma_5 t_i^{f}t^{c}_{j}  \psi_C )(\bar{\psi}_C i \gamma_5
t_i^{f}t^{c}_{j}\psi) +(\bar{\psi} t_i^{f}t^{c}_{j} \psi_C
)(\bar{\psi}_C
t_i^{f}t^{c}_{j}\psi)\right],\label{eqn:Lagrangian2}\\
\mathcal{L}_{\chi}^{(6)}=-K \left\{ \det_{f}\left[ \bar \psi
\left( 1 + \gamma_5 \right) \psi \right] + \det_{f}\left[ \bar
\psi \left( 1 - \gamma_5 \right) \psi \right] \right\},\label{eqn:Lagrangian3}\\
\mathcal{L}_{\chi{d}}^{(6)}=\frac{K'}{8}\sum_{i,j,k=1}^3\sum_{\pm}
\left[({\psi}t_i^{f}t^{c}_k(1 \pm
\gamma_5){\psi}_C)(\bar{\psi}t_j^{f}t^{c}_k(1 \pm
\gamma_5)\bar{\psi}_C)(\bar{\psi}_i( 1 \pm
\gamma_5)\psi_j)\label{eqn:Lagrangian4}\right].
\end{eqnarray}
Here the four-fermion interactions
are all invariant under the $U(3)_R\times{U(3)_L}$-transformation in
the flavor space.  In our notations, the Gell-Mann matrices in
flavor (color) space are $\lambda_i^{f(c)}$ with $i=1,\ldots,8$, and
$\lambda_0^{f(c)}\equiv \sqrt{2/3} \,\openone_{f(c)}$,  and
 the antisymmetric one is denoted by $t_i^{f(c)}$ with $i=1,2,3$ :
\begin{eqnarray}
&t_1^{f(c)}=\lambda_7^{f(c)}, \quad  t_2^{f(c)}=\lambda_5^{f(c)},
\quad t_3^{f(c)}=\lambda_2^{f(c)}.
\end{eqnarray}
The scalar interaction in $\mathcal{L}_{\chi}^{(4)}$ is
responsible for the dynamical chiral symmetry breaking in the
vacuum with the formation of the chiral condensate, while the
vector interaction can be used to investigate the effect
of density-density interaction on the chiral phase
transition ~\cite{Zhang:2009mk}\footnote{We remark that
the effects of the vector interaction on the
baryon-number susceptibility and the chiral transition in the
two-flavor case are examined in \cite{Kunihiro:1991qu} and
\cite{Asakawa:1989bq,Kitazawa:2002bc}, respectively.}.
In Eqs.(\ref{eqn:Lagrangian2}) and (\ref{eqn:Lagrangian4}), $\psi_C$
stands for $C\bar{\psi}^T$ and $C=i\gamma_0\gamma_2$ is the Dirac
charge conjugation matrix. We remark that the suffix $3$ in $t_3^f$ denotes the channel
 for the u$-$d pairing, for example.
For lower temperature $T$ and
large enough baryon chemical potential $\mu$, $\mathcal{L}_{d}^{(4)}$
leads to the formation of diquark condensate in the
color-anti-triplet channel ~\cite{Alford:1997zt, Rapp:1997zu, Alford:1998mk}.
 Besides the four-fermion interactions, Lagrangian (\ref{eqn:Lagrangian})
also contains two types of six-quark interactions,
 $\mathcal{L}_{\chi}^{(6)}$ and $\mathcal{L}_{\chi d}^{(6)}$:
 the former is the traditional Kobayashi-Maskawa-'tHooft (KMT)
interaction ~\cite{Kobayashi:1970ji, 't Hooft:1976fv} and its effect on
the phase diagram in $T$-$\mu$ plane is fully
examined\cite{Kunihiro:1989my,Kunihiro:1991hp,Hatsuda:1994pi,Buballa:2003qv,Fu:2007xc,Kunihiro:2009ds},
whereas the latter could be obtained by a Fierz transformation of the former
and induces the coupling between the chiral and diquark condensates ~\cite{Alford:1998mk,Rapp:1999qa,Steiner:2005jm,Yamamoto:2007ah,Abuki:2010jq}.
We remark that both interactions respect the flavor symmetry
of $SU(3)_R\times SU(3)_L\times U(1)$ while violating the
$U_A(1)$ symmetry as mentioned above. The former is responsible
for accounting for the abnormally large mass of $\eta'$ beyond
the Weinberg inequality\cite{Weinberg:1975ui} (in contrast to
other pseudo Nambu-Goldstone bosons ) in the effective chiral
 model and can be identified as an induced quark interaction
from instantons\cite{'t Hooft:1976fv,ref:SS}. The introduction of the
latter to the Lagrangian expands the study of CSC to the six-fermion
level~\cite{Steiner:2005jm}.

\subsection{The model parameters}

The numerical values of some model parameters are given in
Table~\ref{tab:tab1}. In contrast to~\cite{Abuki:2010jq}
, we only consider the case with realistic quark
masses. The choice of the model parameters is the same as that
in ~\cite{Ruester:2005jc, Zhang:2009mk, Basler:2010xy}
(all following Ref.~\cite{Rehberg:1995kh} ), where $G_S$, the coupling
constant for the scalar meson channel, and $K$, the coupling constant
of the KMT term, are fixed by the vacuum physical observables
(meson masses and decay constants). We shall work in the isospin
symmetric limit in two-flavor space with $m_u=m_d=5.5 \MeV$, and a
sharp three-momentum cut-off $\Lambda$ is adopted.

\begin{table}{
\begin{tabular}{c|c|c|c|c|c}
\hline\hline {\quad $m_{u,d}$[MeV] \quad  } &  \quad {\quad
$m_{s}$[MeV] \quad}&  \quad {\quad $G_S\Lambda^2$ \quad}& \quad
\small {\quad $K\Lambda^5$ \quad}&\quad {\quad $\Lambda$ [MeV]}
\quad &{$M_{u,d}$ [MeV]}\\
\hline 5.5 & 140.7 & 1.835 & 12.36 & 602.3 & 367.7 \\
\hline\hline {\quad $f_{\pi}$[MeV] \quad  } & \quad {\quad
$m_{\pi}$[MeV] \quad} & \quad {\quad $m_K$ [MeV] \quad}& \quad
{\quad $m_{\eta^{,}}$[MeV] \quad} & \quad {\quad $m_{\eta}$[MeV]}
\quad& {$M_{s}$ [MeV]}\\
\hline 92.4 & 135 & 497.7 & 957.8 & 514.8  & 549.5\\
\hline\hline
\end{tabular}
\caption{Model parametrization of two-plus-one-flavor NJL.}
\label{tab:tab1}
}
\end{table}

In contrast to $G_S$ and $K$, no definite observables in the vacuum
are available for determining the coupling constants $G_V, G_D $ and $K'$ in
such a quark model, although we could read off their values from
a Fierz transformation of known vertices:
the coupling constant $K'$, for instance, can be related to $K$
through the Fierz transformation of the instanton vertex, and $K'$
is found to be identical to $K$ ~\cite{Abuki:2010jq}.
Since we are mainly interested in the roles of $K'$ and $G_V$ on
the chiral phase transition and the chromomagnetic instability, both
these coupling constants are treated as free parameters in the
present work. Following \cite{Abuki:2010jq,Basler:2010xy}, we only
consider the attractive interaction between the chiral condensate
and the diquark condensate. Namely, the coupling $K'$ is kept positive.
As for the ratio of $G_D/G_S$, we adopt the standard value from
Fierz transformation in this paper. Due to the contribution
from the KMT interaction, the ratio $G_D/G_S$ from Fierz transformation
should be 0.95 rather than 0.75 obtained by only considering the four-quark
interaction ~\cite{Buballa:2003qv}. Such a choice of the coupling has also been
used in Refs.~\cite{Zhang:2009mk} and ~\cite{Basler:2010xy}. In the literature,
the diquark-diquark interaction near the standard value from Fierz
transformation is usually called the intermediate coupling.

\subsection{Thermodynamic potential with the constraints of charge-neutrality and $\beta$-equilibrium}

The grand partition function of the NJL model is given by
\begin{equation}
Z\equiv{e^{-{\Omega}V/T}}=\int{D\bar{\psi}D\psi}
e^{i\int{dx^4}(\cal{L}+{\psi^\dag}\hat{\mu}\psi)},
\label{eqn:gpf}
\end{equation}
where $\Omega$ is the thermodynamic potential density and
$\hat{\mu}$ is the quark chemical potential matrix. In general,
the quark chemical potential matrix $\hat{\mu}$ takes the
form~\cite{Alford:2002kj}
\begin{equation}
 \hat{\mu} = \mu - \mu_e Q + \mu_3T_3 + \mu_8T_8, \label{eqn:chemicalp}
\end{equation}
where $\mu$ is the quark chemical potential (i.e.\ one third of
the baryon chemical potential), $\mu_e$ the chemical potential
associated with the (negative) electric charge, and $\mu_3$ and
$\mu_8$ represent the color chemical potentials corresponding to
the Cartan subalgebra in the SU(3)-color space. The explicit form
of the electric charge matrix is
$Q=\text{diag}(\frac{2}{3},-\frac{1}{3},-\frac{1}{3}))$ in flavor
space, and the color charge matrices are
$T_3=\text{diag}(\frac{1}{2},-\frac{1}{2},0)$ and
$T_8=\text{diag}(\frac{1}{3},\frac{1}{3}, -\frac{2}{3})$ in the color
space.  The chemical potentials for the quarks with
respective flavor and color charges are listed
below:
\begin{equation}
 \begin{split}
&\mu_{ru} = \mu-\tfrac{2}{3}\mu_e+\tfrac{1}{2}\mu_3+\tfrac{1}{3}\mu_8 \,,
 \quad
 \mu_{rd} = \mu+\tfrac{1}{3}\mu_e+\tfrac{1}{2}\mu_3+\tfrac{1}{3}\mu_8 \,,
 \quad
 \mu_{rs} = \mu+\tfrac{1}{3}\mu_e+\tfrac{1}{2}\mu_3+\tfrac{1}{3}\mu_8 \,,\\
&\mu_{gu} = \mu-\tfrac{2}{3}\mu_e-\tfrac{1}{2}\mu_3+\tfrac{1}{3}\mu_8 \,,
 \quad
 \mu_{gd} = \mu+\tfrac{1}{3}\mu_e-\tfrac{1}{2}\mu_3+\tfrac{1}{3}\mu_8 \,,
 \quad
 \mu_{gs} = \mu+\tfrac{1}{3}\mu_e-\tfrac{1}{2}\mu_3+\tfrac{1}{3}\mu_8 \,,\\
&\mu_{bu} = \mu-\tfrac{2}{3}\mu_e-\tfrac{2}{3}\mu_8 \,, \qquad\qquad
 \mu_{bd} = \mu+\tfrac{1}{3}\mu_e-\tfrac{2}{3}\mu_8 \,, \qquad\qquad
 \mu_{bs} = \mu+\tfrac{1}{3}\mu_e-\tfrac{2}{3}\mu_8 \,.
 \end{split}
\end{equation}

Corresponding to the chiral and diquark interactions in
Eq.~(\ref{eqn:Lagrangian}), we assume that the following condensates
are formed in the system, namely, the scalar quark-antiquark condensate
\begin{equation}
\sigma_i = \langle \bar
\psi_i\psi_i \rangle,\label{eqn:chiralcon}
\end{equation}
and the scalar diquark condensate
\begin{equation}
s_i =
\langle \bar{\psi}_Ci\gamma_5 t_i^ft_i^c \psi \rangle\label{eqn:diqcon}.
\end{equation}
In addition, we remark that the quark-number (or baryon-number) density
\begin{equation}
\rho_i =
\langle \bar \psi_i\gamma^{0}\psi_i \rangle,\label{eqn:vectorcon}
\end{equation}
has a finite value for finite $\mu$. Note that the indices 1,2 and 3 in
Eqs.~(\ref{eqn:chiralcon}) and (\ref{eqn:vectorcon}) represent u, d
and s quarks, respectively, whereas in Eq.~(\ref{eqn:diqcon}), the indices
 1, 2 and 3 stand for the diquark condensate in d-s, s-u and u-d pairing
channels, respectively. Here we have assumed the condensates and
the density are all homogeneous; the study of the phase structure with
inhomogeneous condensates and/or baryon-number density
\cite{Nakano:2004cd,Nickel:2009ke,Nickel:2009wj,Carignano:2010ac,Giannakis:2004pf}
 is surely intriguing but beyond the scope of the present work.

The constituent quark masses and the dynamical Majarona masses are
expressed in terms of these condensates as follows:
\begin{equation}
 M_i =  m_i - 4G_S \sigma_i+K|\varepsilon_{ijk}|\sigma_j\sigma_k+
 \frac{K'}{4}|s_i|^2\,,\label{eqn:cmass}
\end{equation}
and
\begin{equation}
\Delta_i=2(G_D-\frac{K'}{4}\sigma_i)s_i\,.\label{eqn:mmass}
\end{equation}
Similarly, it is convenient to define
 the dynamical quark chemical potential for flavor $i$ by
\begin{equation}
{\tilde{\mu}}_i = \mu_i-4G_V{\rho_i}\,.\label{eqn:dchept}
\end{equation}

A few remarks are in order here:
\begin{enumerate}
\item ~Both types of the anomaly terms ${\mathcal{L}_{\chi}}^{(6)}$ and
$\mathcal{L}_{\chi{d}}^{(6)}$
contribute to the constituent quark
masses in Eq.~(\ref{eqn:cmass}), and thus if $K'$ and the diquark condensate $s_i$ are finite,
chiral symmetry is dynamically broken even when the usual chiral condensates are absent.\,
\item~The new anomaly term $\mathcal{L}_{\chi{d}}^{(6)}$
also modifies the formula for the Majarona mass for the CSC
phase so that the chiral condensates affects the Majorana mass, and
hence induce an interplay between the two condensates:
Indeed the `bare' diquark-diquark coupling $G_D$ is replaced by
an effective one, ${G_D'}_i\equiv G_D-\frac{K'}{4}\sigma_i$,
as shown in Eq.~(\ref{eqn:mmass}),
which is dependent on the chiral condensates.
 Thus the flavor-dependent effective coupling ${G_D'}_i$ is now
dependent on $T$ and $\mu$ through $\sigma_i$.\,
\item~Equations.~(\ref{eqn:cmass}) and~(\ref{eqn:mmass})
clearly show that  the flavor-mixing
occurs not only in the usual chiral condensates due to ${\mathcal L}_{\chi}^{(6)}$ but also
in the diquark condensates owing to ${\mathcal L}_{\chi d}^{(6)}$, which would
lead to interesting  physical consequences.\,
\item~It is also to be noted that the dynamical quark chemical potential $\tilde{\mu}_i$
for u and d quarks are different from each other because of the
constraint of electric charge-neutrality ($\mu_d>\mu_u$) in 2CSC;
notice also, however, that they are dependent only on the
respective  density $\rho_{u,d}$ and hence the dynamical chemical
potentials $\tilde{\mu}_{u, d}$ tend to come closer because
$\rho_d>\rho_u$ with the common coupling constant $G_V$ \cite{Zhang:2009mk}.
\end{enumerate}

In the mean field level, the thermodynamic potential for the
two-plus-one-flavor NJL  with the charge-neutrality constraints reads
\begin{eqnarray}
\Omega &=& \Omega_{l} +2 G_S \sum_{i=1}^{3} \sigma_i^2 -2 G_V \sum_{i=1}^{3} \rho_i^2
+ \sum_{i=1}^{3}(G_D-\frac{K'}{2}\sigma_i)\left|s_i \right|^2\nonumber\\
&-& 4 K \sigma_1 \sigma_2 \sigma_3 -\frac{T}{2V} \sum_P \ln \det
\frac{S^{-1}_{MF}}{T} \;.
\label{eqn:Omega2}
\end{eqnarray}
Notice the presence of the new cubic-mixing terms among the chiral
and diquark condensates. In Eq.~(\ref{eqn:Omega2}), $\Omega_{l}$ denotes
the contribution from free leptons. Note that $\Omega_{l}$ should include the contributions from
both electrons and muons  for completeness. Since $M_\mu>>M_e$
and $M_e\approx 0$, ignoring the contribution of muons has little
effect on the phase structure. Therefore, only electrons are considered
in our calculation and the corresponding $\Omega_{l}$ reads
\begin{equation}
\Omega_{l}=-\frac{1}{12\pi^2}\left(\mu_e^4+2\pi^2T^2\mu_e^2
            +\frac{7\pi^4}{15}T^4\right) \,\label{eqn:lepton}.
\end{equation}

Due to the large mass difference between s and u [d]
quarks, the most favored phase at low temperature and  moderate density
tends to be the 2CSC  rather than CFL phase, as demonstrated in
the two-plus-one-flavor NJL model~\cite{Ruester:2005jc,Abuki:2005ms}.
Surprisingly enough, if the anomaly term ${{\mathcal L}_{\chi d}}^{(6)}$
is incorporated, the 2CSC phase turns to be still favored
in the intermediate density region even when the three flavors
have the equal mass~\cite{Basler:2010xy}. Needless to say, the
dominance of the 2CSC phase over the CFL one is more robust
when the realistic mass hierarchy for the three
flavors is adopted. Moreover, it is worth mentioning here that
the mass disparity favors the 2CSC phase with the u-d pairing
also through the inequality of the effective diquark coupling
${G_D'}_3>{G_D'}_{1,2}$ when the anomaly coupling $K'$ is present;
see  Eq.~(\ref{eqn:mmass}). Since the main purpose of the present work
is to explore how the axial anomaly term ${\mathcal L}_{\chi d}^{(6)}$ affect
the phase boundary involving the chiral transition
at moderate densities, under the constraints of
the charge-neutrality and $\beta$-equilibrium, we only consider
the 2CSC phase in the following. Note that
$\mu_3$ in (\ref{eqn:chemicalp}) vanishes in the 2CSC phase
because the color SU(2) symmetry for the red and green quarks
are left unbroken.

\begin{figure}
\hspace{-.0\textwidth}
\begin{minipage}[t]{.5\textwidth}
\includegraphics*[width=\textwidth]{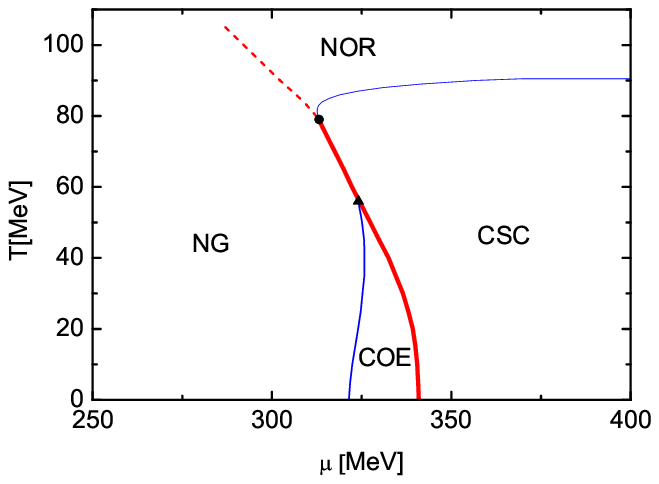}
\centerline{(a) $K'/K=2.0$}
\end{minipage}
\hspace{-.05\textwidth}
\begin{minipage}[t]{.5\textwidth}
\includegraphics*[width=\textwidth]{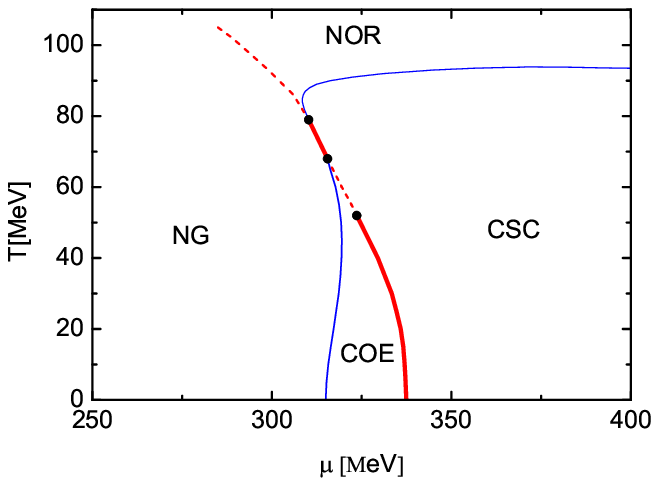}
\centerline{(b) $K'/K=2.25$}
\end{minipage}
\hspace{-.0\textwidth}
\begin{minipage}[t]{.5\textwidth}
\includegraphics*[width=\textwidth]{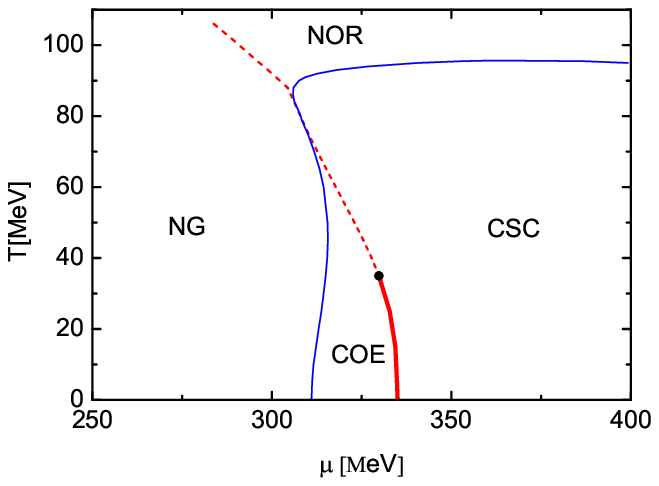}
\centerline{(c) $K'/K=2.4$}
\end{minipage}
\hspace{-.05\textwidth}
\begin{minipage}[t]{.5\textwidth}
\includegraphics*[width=\textwidth]{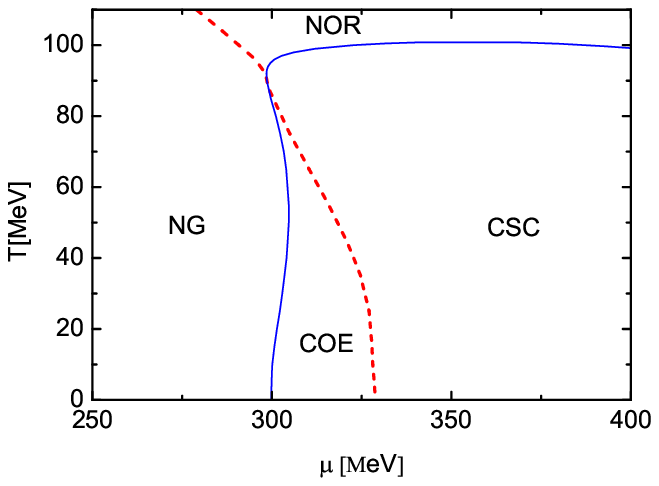}
\centerline{(d) $K'/K=2.8$}
\end{minipage}
\hspace{.0\textwidth} \caption{The phase diagrams in
the $T$-$\mu$ plane for various values
of  $K'$ in the two-plus-one-flavor NJL model with the charge-neutrality and
$\beta$-equilibrium being kept. The vector interaction
is not taken into account. The thick solid line, thin solid line
and dashed line denote the first order transition, second order transition
and chiral crossover, respectively. }
\label{fig: pdzeroGv}
\end{figure}

The inverse quark-propagator matrix in
the Nambu-Gor'kov formalism takes the following form in the
mean-field approximation,
 \begin{equation}
  S^{-1}_{\mathrm{MF}}(i\omega_n,\vec{p}) = \bigg(\begin{array}{cc}
  [{G_0^{+}}]^{-1} & \Delta\gamma_5t_3^ft_3^c \\
  -\Delta^*\gamma_5t_3^ft_3^c &
  [{G_0^{-}}]^{-1} \end{array}\bigg) \,,
\end{equation}
with
\begin{equation}
 [{G_0^{\pm}}]^{-1}=\gamma_0(i\omega_n\pm\hat{\tilde{\mu}})
  -\vec{\gamma}\cdot\vec{p}-\hat{M} \, ,
\end{equation}
where $\hat{M}={\rm diag}_f(M_u,M_d,M_s)$,
$\hat{\tilde{\mu}}={\rm diag}_f(\tilde{\mu}_u,\tilde{\mu}_d,\tilde{\mu}_s)$
and $\omega_n=(2n+1)\pi{T}$ is the Matsubara frequency.
Taking the Matsubara sum, the last part of the thermodynamic
potential~(\ref{eqn:Omega2}) is expressed as
\begin{equation}
 -\frac{T}{2V} \sum_P \ln \det
\frac{S^{-1}_{MF}}{T}=-\sum_{i=1}^{18}\int\frac{d^3p}{(2\pi)^3}\{(E_i-E_i^0)+2T\ln(1+e^{-E_i/T})\},
\label{eqn:therp}
\end{equation}
with the dispersion relations for
nine quasi-particles (that is, three
flavors $\times$ three colors;  the spin degeneracy is already taken
into account in Eq.~(\ref{eqn:therp})) and
nine quasi-antiparticles. In Eq.~(\ref{eqn:therp}),
$E_i^0$ represents $E_i(M=m,\Delta=0,\rho=0)$. The s quark and unpaired
blue u and d quarks have twelve energy dispersion relations with
a similar form. For example, the dispersion relations for the blue u quark
and anti blue u quark are
\begin{equation}
 E_{bu} = E - \tilde{\mu}_{bu}\quad \text{and} \quad \bar{E}_{bu} = E + \tilde{\mu}_{bu} \,,
\end{equation}
respectively, with $E=\sqrt{\vec{p}^2+{M_u^2}}$.  In the $rd$-$gu$ quark sector
with pairing we can find the four dispersion relations,
\begin{equation}
 \begin{split}
 E_{\text{$rd$-$gu$}}^{\pm} = E_\Delta \pm \tfrac{1}{2}(\tilde{\mu}_{rd}-\tilde{\mu}_{gu})
  = E_\Delta \pm \delta\tilde{\mu}\,,\\
 \bar{E}_{\text{$rd$-$gu$}}^{\pm} = \bar{E}_\Delta \pm
  \tfrac{1}{2}(\tilde{\mu}_{rd}-\tilde{\mu}_{gu})
  = \bar{E}_\Delta \pm \delta\tilde{\mu} \,,
 \end{split}
\end{equation}
and the $ru$-$gd$ sector has another four as
\begin{equation}
 \begin{split}
 E_{\text{$ru$-$gd$}}^{\pm} = E_\Delta \pm \tfrac{1}{2}(\tilde{\mu}_{ru}-\tilde{\mu}_{gd})
  = E_\Delta \mp \delta\tilde{\mu} \,,\\
 \bar{E}_{\text{$ru$-$gd$}}^{\pm} = \bar{E}_\Delta \pm
  \tfrac{1}{2}(\tilde{\mu}_{ru}-\tilde{\mu}_{gd})
  = \bar{E}_\Delta \mp \delta\tilde{\mu} \,,
 \end{split}
\end{equation}
where $E_\Delta=\sqrt{(E-\bar{\tilde{\mu}})^2+\Delta^2}$ and
$\bar{E}_\Delta=\sqrt{(E+\bar{\tilde{\mu}})^2+\Delta^2}$;
from now on $\Delta$ stands for $\Delta_3$.
The average chemical potential is defined by
\begin{equation}
 \bar{\tilde{\mu}} = \frac{\tilde{\mu}_{rd}+\tilde{\mu}_{gu}}{2}
 = \frac{\tilde{\mu}_{ru}+\tilde{\mu}_{gd}}{2} = \mu-\frac{\mu_e}{6}
 -2G_V(\rho_1+\rho_2)+ \frac{\mu_8}{3} \,,
\label{Average}
\end{equation}
and the effective mismatch between the chemical potentials of u
and d quarks takes the form
\begin{equation}
\delta\tilde{\mu}=\tfrac{1}{2}(\mu_e-4G_V(\rho_2-\rho_1)).
\label{eqn:Mismatch}
\end{equation}

Ignoring the the mass difference between u and d quarks, the
determinantal term in Eq.~(\ref{eqn:Omega2}) has an
analytical form which greatly simplifies the
numerical calculation. Adopting the variational method, we get the
eight non-linear coupling equations
\begin{equation}
 \frac{\partial\Omega}{\partial\sigma_1}=
 \frac{\partial\Omega}{\partial\sigma_3}=
 \frac{\partial\Omega}{\partial {s_3}}=
 \frac{\partial\Omega}{\partial\rho_{1}}=
 \frac{\partial\Omega}{\partial\rho_{2}}=
\frac{\partial\Omega}{\partial\rho_{3}}=
 \frac{\partial\Omega}{\partial\mu_e}=
 \frac{\partial\Omega}{\partial{\mu_8}}=0\,.
 \label{eqn:gapeq}
\end{equation}
Since $\mu_8$ is tiny  around the chiral transition region
\cite{Ruester:2005jc,Abuki:2005ms}, we shall set it zero
with little effect in the numerical results~\cite{Zhang:2009mk}.
Thus, Eq.~(\ref{eqn:gapeq}) is then simplified to
a system of seven coupled equations.

\begin{figure}
\hspace{-.05\textwidth}
\includegraphics*[width=0.9\textwidth]{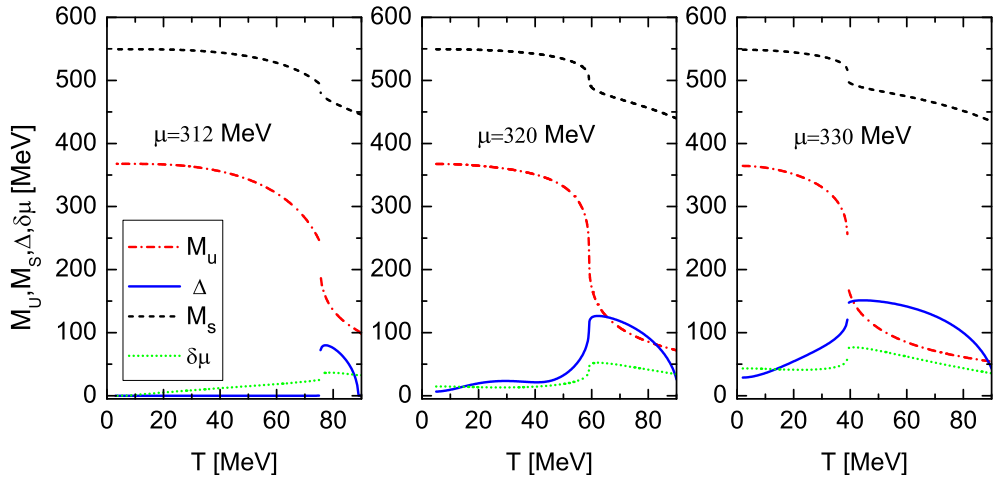}
\caption{ The temperature dependence of $M_u, M_s , \Delta$ and
$\delta\tilde{\mu}$ for fixed  $K'/K=2.25$  and  three different
chemical potentials $\mu=312$ MeV, 320 MeV and 330 MeV. The
constraints of electric charge-neutrality and $\beta$-equilibrium
are imposed while the vector interaction is not incorporated. }
\label{fig:abnormal1}
\end{figure}

\section{Phase Structure With The Axial Anomaly}

In this section, we show numerical results
of the effects of the new six-quark
interaction (\ref{eqn:Lagrangian4}) on the chiral phase transition
under the charge-neutrality and $\beta$-equilibrium constraints
with or without the vector interaction. Since we are mainly interested
in the phase diagram involving chiral transition at low temperatures,
all the phase diagrams will be plotted for the range
$250\text{MeV}<\mu<400 \text{MeV}$ where the chiral transition is
expected to be relevant. As for the type of the CSC phase,
Ref.~\cite{Basler:2010xy} indicates that the CFL phase is only realized
for $\mu>460\text{MeV}$ even when $K'=0$ with the same model parameters as ours,
and an increase of $K'$ pushes the CFL phase to even higher $\mu$ region.
Therefore, we exclusively consider the 2CSC phase
near the chiral boundary without a loss of generality.

In the following, we use the same notations as in
Ref.~\cite{Hatsuda:2006ps,Zhang:2008wx, Zhang:2009mk} to distinguish
the different regions in the $T$-$\mu$ phase diagram. Namely,
NG, CSC, COE, and NOR refer to the hadronic (Nambu-Goldstone) phase with
$\sigma\neq0$ and $\Delta=0$, color-superconducting phase with
$\Delta\neq0$ and $\sigma=0$, coexisting phase with $\sigma\neq0$
and $\Delta\neq0$, and normal phase with $\sigma=\Delta=0$,
respectively, though they have exact meanings only in the chiral
limit.

\subsection{The case without vector interaction}

We first show the numerical results in the case without
the vector interaction. The phase diagrams with varying coupling
constant $K'$ are displayed in Fig.~\ref{fig: pdzeroGv}.
In contrast to Fig.~8 in Ref.~\cite{Basler:2010xy},
the multi-CP structure can still appear with a choice of $K'$
in the phase diagram when the charge-neutrality, $\beta$-equilibrium and
the new axial anomaly term are simultaneously taken into account.
For $K'/K=2.0$, Fig.~\ref{fig: pdzeroGv}a
shows that there exists only one usual chiral CP even though the COE
emerges: We remark that the COE region does not exist when $K'/K=0$, which
is not displayed in Fig.~\ref{fig: pdzeroGv}. Figure~\ref{fig: pdzeroGv}b
shows that when $K'/K$ is increased to 2.25,
the chiral transition turns to a crossover at relatively lower
temperatures, and hence there appear two new chiral CP's.
With a further increase of $K'/K$, the boundary line for first-order
transition at higher temperature shrinks and thus
the two crossover boundary lines in
Fig.~\ref{fig: pdzeroGv}b join with each other, and eventually only one
CP is left in the phase diagram, as shown in Fig.~\ref{fig: pdzeroGv}c.
When $K'/K$ is large enough, Fig.~\ref{fig: pdzeroGv}d indicates that the
first order boundary vanishes completely and there is no chiral CP in the
phase diagram.

We note that the emergence of the three CP's in
Fig.~\ref{fig: pdzeroGv}b comes from a joint effect of the interplay
between the chiral and diquark condensates and the electric
charge-neutrality constraint. First of all, we recall that
the abnormal thermal behavior of the diquark condensate that it has
a maximum at a finite temperature in the COE
is responsible for the emergence of the multiple chiral CP structure
\cite{Kitazawa:2002bc,Addenda,Zhang:2008wx,Zhang:2009mk}.
Such a behavior is also observed in the present case,
as displayed in Fig.~\ref{fig: pdzeroGv}b. As first indicated in
Ref.~\cite{Zhang:2008wx}, when $\mu_e=\mu_d-\mu_u$  is positive, the
boundary of the chiral transition is shifted towards higher $\mu$
region, and leads to the formation of the COE at low-temperature
region, in which the chiral phase transition is significantly
weakened by the smearing of the Fermi surface inherent in the
CSC phase. In this regard, $\mu_e$
plays a role of an effective vector
interaction~\cite{Kitazawa:2002bc,Addenda,Zhang:2009mk}.
On the other hand, the chiral anomaly term with
positive $K'$ intensifies the competition between the
chiral and diquark condensates due to the enhanced
effective diquark-diquark interaction. Thus when $K'$ is increased,
the CSC region expands towards lower $\mu$ region in the $T$-$\mu$ plane.
Consequently, the COE region tends to be more easily formed
when both $\mu_e$ and $K'$ take effects. Therefore the chiral transition
is significantly weakened and the smooth crossover gets to appear
with new CP's in the intermediate temperature owing to the abnormal
thermal behavior of the diquark condensate.

The $T$-dependence of $M_u, M_s, \Delta$ and $\delta\tilde{\mu}$ for
fixed $K'/K=2.25$ and several values of $\mu$ is shown in Fig.~\ref{fig:abnormal1}.
One can see that, with increasing $T$, the constituent quark masses
decrease persistently while the Majarona mass for CSC first increases,
has the maximum value and then decreases in the COE region and nearby.
Let us see the details for each value of $\mu$.
For a small $\mu=312$ MeV, the diquark pairing is weak and the
gap $\Delta$ does not appear at lower temperature region.
Thus the chiral phase transition keeps the nature of the first order
at $T_{C1}\approx{75}$ MeV. At a lager $\mu=320$ MeV, the diquark pairing
becomes significant and the $\Delta$ shows the abnormal thermal
behavior with a maximum value around $T_{C2}\approx60$\, MeV, and hence the chiral
phase transition turns to a smooth crossover owing to the competition
with the diquark condensate in the COE region. For even larger
 $\mu=330$ MeV, the diquark pairing becomes more significant and
the $\Delta$ still shows the abnormal thermal behavior.
However, the competition between the two condensates is not strong
enough to qualitatively change the nature of the chiral restoration
and a first order transition happens at $T_{C3}\approx40$ MeV.
The reason why the crossover does not occur at $\mu=330$ MeV but
happens at $\mu=320$ MeV can be understood as follows:
Starting from the same point ($T=T_{C3}$, $\mu=320$ MeV) in the COE region,
an increase of $T$ affects the nature of the chiral transition more
significantly than that of $\mu$ does, since $T_{C3}<T_{C2}$.

We have seen that the abnormal thermal behavior of the
gap $\Delta$ plays an essential role in realizing the
multi-CP structure of the phase diagram.
Such an unusual $T$-dependence of the $\Delta$ can be
attributed to the following two mechanisms:
(i)~the mismatch between the chemical potentials
of u  and d quarks owing to the charge-neutrality and
$\beta$-equilibrium constraints and
(ii)~the small Fermi spheres of the quarks in the COE region
due to the relatively large quark masses:
First, the difference in the chemical potentials
$\delta\tilde{\mu}$ or the mismatch of the Fermi momenta
disfavors the u-d pairing at zero or small temperature.
However, as the temperature is raised in the low-$T$ region,
more and more u and d quarks tends to participate in the
pairing due to the smearing of the Fermi surfaces, especially
that of the u quark. Of course, when $T$ is raised too much,
the pairing will be gradually destroyed.
Thus the $\Delta$ will have the maximum value at a finite $T$ and
then disappears eventually when $T$ is further raised.
These dual effects of the temperature
on the diquark pairing lead to the abnormal behavior of the $\Delta$.
This behavior becomes more prominent in the 2CSC phase for a
weak diquark coupling \cite{Shovkovy:2004me} or in the COE region for
a moderate or strong diquark coupling \cite{Zhang:2008wx,Zhang:2009mk}.
Second, when $T$ is raised, the dynamical quark masses decrease
and hence the Fermi spheres or momenta of u and d quarks grow
significantly for a fixed $\mu$, which means that the density of
states at the Fermi surface increases with $T$, and thus the diquark
pairing is enhanced in the COE region. The increased diquark condensates
in turn tend to further suppress the dynamical quark masses.

Notice that such an increase of the diquark condensate along with increasing $T$
is expected to be most prominent around the phase boundary of the chiral
transition, including the COE region, where the chiral condensates change
most significantly. For the neutral 2CSC, once the COE is formed, both of these
mechanisms take effects simultaneously and are mutually enhanced,
and thus the formation of the multiple-CP structure is readily made.

This may explain why no intermediate-temperature CP is realized in
Ref.~\cite{Basler:2010xy} where the chiral-diquark interplay is
embodied by the anomaly term but without the charge-neutrality and
$\beta$-equilibrium constraints; the anomaly term solely is insufficient for
realizing the abnormal thermal behavior of the $\Delta$.
It should be stressed that the mechanism for the
emergence of the intermediate-temperature CP's in Fig.~\ref{fig: pdzeroGv}b is
apparently similar to that in the two-flavor case found in ~\cite{Zhang:2008wx}.
However, the strange quark plays an important role in the present case
since the chiral condensate of the strange quark
contributes positively to the effective diquark-diquark coupling for
u and d quarks through the axial anomaly.
We should stress that apart from the appearance of intermediate-temperature CP's,
there is a common feature with and without charge-neutrality constraint:
the chiral transition in the low-$T$ region extending zero temperature keeps
first order provided that $K'$ does not exceed a critical value at which
the first-order line completely disappears.

Last but not least, we remark that the $T$-$\mu$ region where
the two new low-temperature CP's are located in Fig.~\ref{fig: pdzeroGv}b
is free from the chromomagnetic instability, which is obvious from
Fig.~\ref{fig:unstable-stable}a; a detailed discussion on this point
will be given in Sec.IV.

\subsection{The case for nonzero vector interaction}

In this subsection, we will investigate the phase diagram when
both the vector and the new six-quark interactions are present
under the charge-neutrality and $\beta$-equilibrium constraints.

There are some choices for the value of the vector coupling:
The chiral instanton-anti-instanton molecule model ~\cite{ref:SS}
gives the ratio $G_V/G_S= 0.25$,
while the Fierz transformation of the vertex given in the truncated
Dyson-Schwinger  model ~\cite{ref:RWP} gives the ratio 0.5.
Thus we rather treat the $G_V/G_S$ as a free parameter
in the range, $0$-$0.5$.

\begin{figure}
\hspace{-.0\textwidth}
\begin{minipage}[t]{.5\textwidth}
\includegraphics*[width=\textwidth]{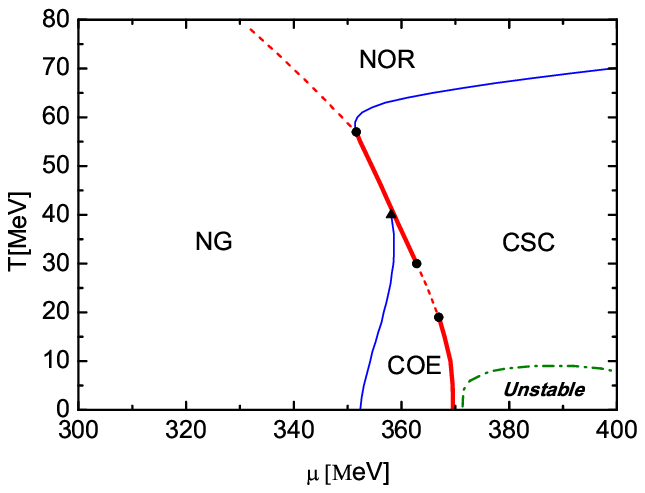}
\centerline{(a) $K'/K=0.55$}
\end{minipage}
\hspace{-.05\textwidth}
\begin{minipage}[t]{.5\textwidth}
\includegraphics*[width=\textwidth]{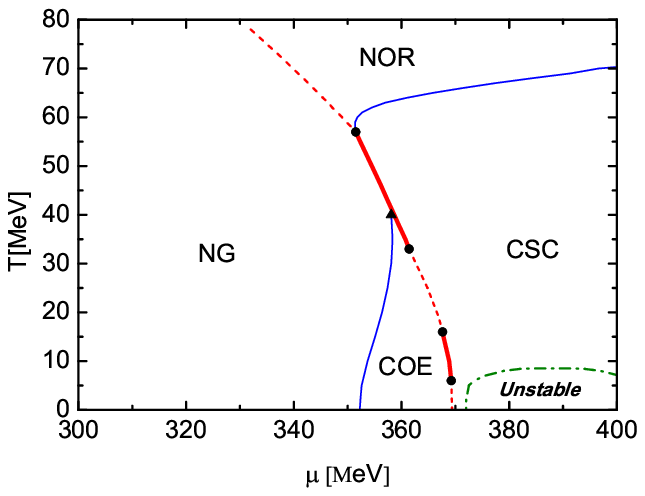}
\centerline{(b) $K'/K=0.57$}
\end{minipage}
\hspace{-.05\textwidth}
\begin{minipage}[t]{.5\textwidth}
\includegraphics*[width=\textwidth]{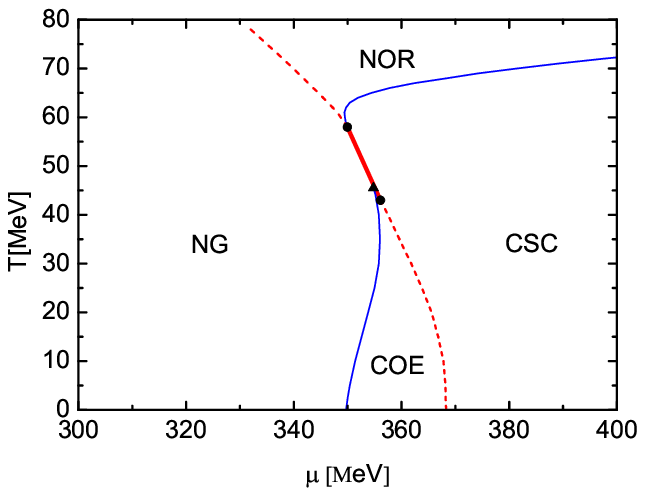}
\centerline{(c) $K'/K=0.70$}
\end{minipage}
\hspace{-.05\textwidth}
\begin{minipage}[t]{.5\textwidth}
\includegraphics*[width=\textwidth]{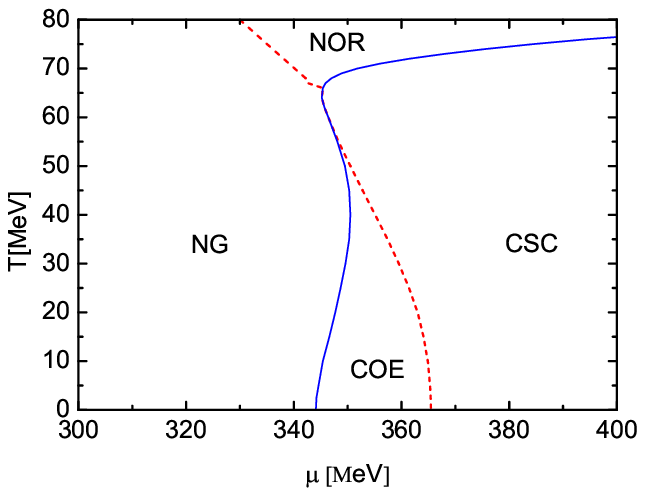}
\centerline{(d) $K'/K=1.0$}
\end{minipage}
\caption{The $T$-$\mu$ phase diagrams of the two-plus-one-flavor NJL model
for several values of $K'/K$ and fixed $G_V/G_S=0.25$, where
the charge-neutrality constraint and $\beta$-equilibrium condition
are imposed. With the increase of $K'/K$, the number of the
critical points changes and the unstable region characterized
by the chromomagnetic instability (bordered by the dash dotted line)
tends to shrink and ultimately vanishes in the phase diagram.
The respective meanings of the various types of lines are
 the same as those in Fig. 1.}
\label{fig:pdGVfixed}
\end{figure}

We first explore the phase diagram in the $T$-$\mu$ plane by varying the
ratio $K'/K$ but with $G_V/G_S$ being fixed as 0.25, the value
given in the instanton-anti-instanton molecule model. When $K'/K$
is small and less than $0.5$, only the usual phase structure with
single CP is obtained.  When $K'/K$ exceeds $0.5$, other four different
types of the CP structures appear, as displayed in Fig.~\ref{fig:pdGVfixed}.
At $K'/K=0.55$, a phase diagram similar to that in Fig.~\ref{fig: pdzeroGv}b
is obtained, as shown in Fig.~\ref{fig:pdGVfixed}a,
where two new intermediate-temperature CP's emerge. When $K'/K$ is slightly
increased to 0.57, the chiral transition becomes crossover
in the lower-temperature region which extends to zero temperature;
thus the total number of the
CP's becomes four, which indicates stronger competition between
the chiral and  diquark condensates at relatively larger $\mu$.
Further increasing $K'/K$, the low-temperature chiral boundary
totally turns into a crossover one and only one first-order
transition line with two CP's attached remains in the phase
diagram, as displayed in Fig.~\ref{fig:pdGVfixed}c. In this case, the
number of the CP's is reduced to two accordingly. When $K'/K$ is large enough,
Fig.~\ref{fig:pdGVfixed}d shows that only chiral crossover
transition exists in the phase diagram with no CP.

In comparison with Fig.~\ref{fig: pdzeroGv} where the vector
interaction is not included,  Fig.~\ref{fig:pdGVfixed} indicates
that the phase structures with multiple CP's can be realized
with relatively small $K'$ owing to the vector interaction.  We remark
that all the types of the chiral CP structures displayed in
Fig.~\ref{fig:pdGVfixed} by varying $K'$ are obtained by varying $G_V$
without the anomaly term~\cite{Zhang:2009mk}.
In the present case,  the number of the critical points changes as
 1$\rightarrow$\, 3\,$\rightarrow$\,4\,$\rightarrow$\,2\,$\rightarrow$\,0 when
$K'$ is increased.

\begin{figure}
\hspace{-.0\textwidth}
\begin{minipage}[t]{.5\textwidth}
\includegraphics*[width=\textwidth]{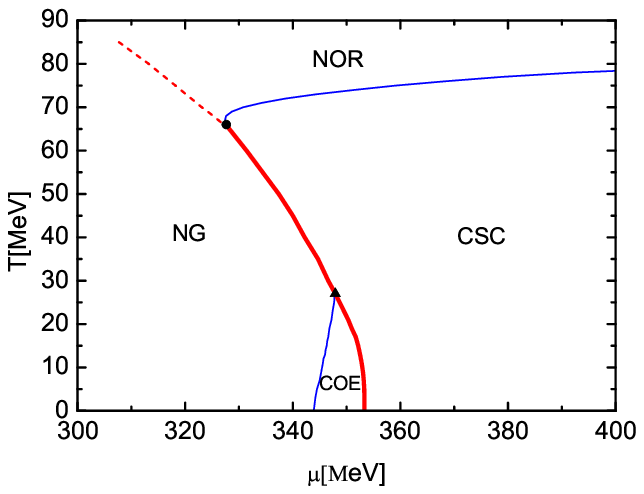}
\centerline{(a) $G_V/G_S=0$}
\end{minipage}
\hspace{-.05\textwidth}
\begin{minipage}[t]{.5\textwidth}
\includegraphics*[width=\textwidth]{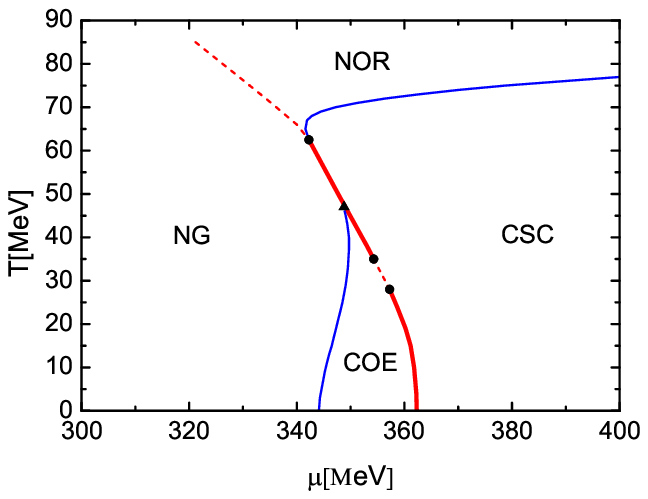}
\centerline{(b) $G_V/G_S=0.193$}
\end{minipage}
\hspace{-.05\textwidth}
\begin{minipage}[t]{.5\textwidth}
\includegraphics*[width=\textwidth]{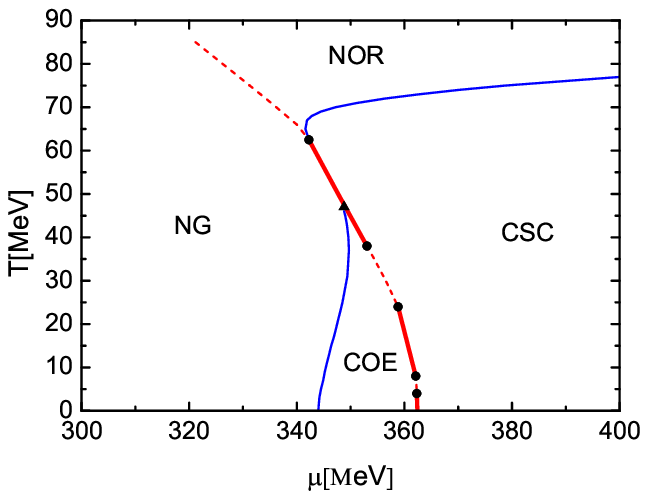}
\centerline{(c) $G_V/G_S=0.195$}
\end{minipage}
\hspace{-.05\textwidth}
\begin{minipage}[t]{.5\textwidth}
\includegraphics*[width=\textwidth]{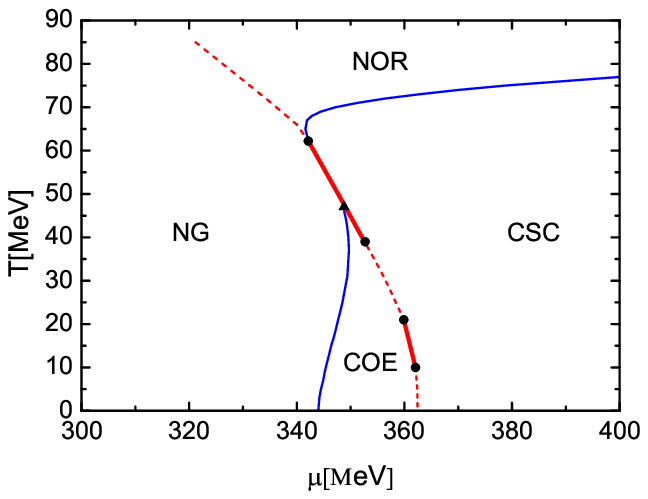}
\centerline{(d) $G_V/G_S=0.197$}
\end{minipage}
\hspace{-.05\textwidth}
\begin{minipage}[t]{.5\textwidth}
\includegraphics*[width=\textwidth]{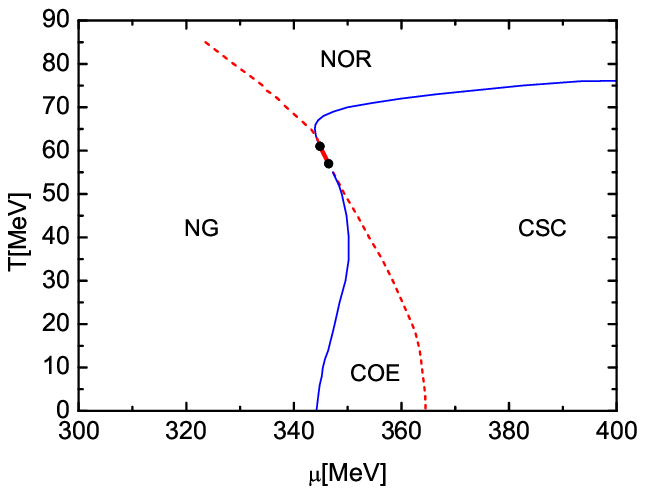}
\centerline{(e) $G_V/G_S=0.23$}
\end{minipage}
\hspace{-.05\textwidth}
\begin{minipage}[t]{.5\textwidth}
\includegraphics*[width=\textwidth]{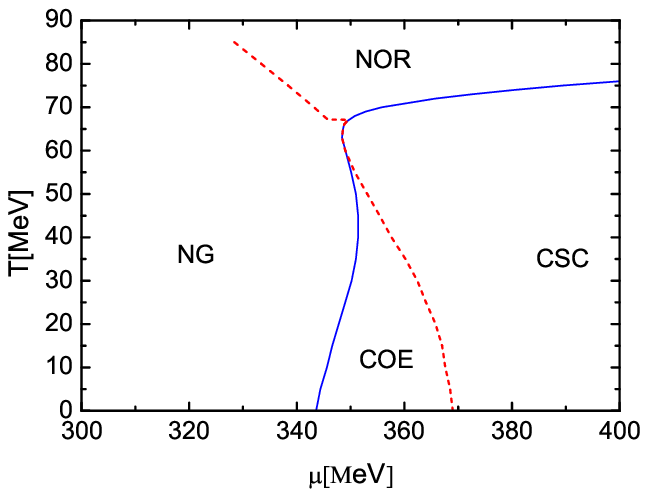}
\centerline{(f) $G_V/G_S=0.3$}
\end{minipage}
\caption{
The phase diagrams in the two-plus-one-flavor NJL model for
fixed $K'/K=1.0$ with $G_V/G_S$ being varied, where the
charge-neutrality constraint and $\beta$-equilibrium condition
are taken into account. The meanings
of the different line types are the same as those in Fig. 1.
The number of the critical points changes along with an
increase of $G_V/G_S$. All the phase
diagrams are free from the chromomagnetic instability.
}
\label{fig:pdK'fixed}
\end{figure}

As is mentioned before, the Fierz transformation of the instanton vertex
leads to the identity $K'=K$, so it is of special interest to investigate
the phase diagram in the case of $K'=K$. A series of phase diagram with
fixed $K'/K=1$ but varied $G_V$ are shown in Fig.~\ref{fig:pdK'fixed}.
One finds that all the chiral CP structures in Fig.~\ref{fig:pdGVfixed}
still appear in the phase diagrams, and moreover,
even as large as five CP's can exist in the phase diagram, as shown
in Fig.~\ref{fig:pdK'fixed}c. This suggests that the interplay between
the chiral and the diquark condensates in the COE region becomes
complicated once the charge-neutrality, the vector interaction and
the axial anomaly are all taken into account. A comparison with the case
of vanishing $K'$, which is given in  Fig.~7 in ~\cite{Zhang:2009mk}, shows
that the parameter range of  $G_V$ for realizing the low-temperature CP's
moves towards lower $G_V$, which is actually natural because
$K'$ gives the same effect as $G_V$ on the chiral transition.
It is noteworthy that such lower values of $G_V$
are also close to the standard value of $G_V/G_S$ derived from
the instanton model. The number of the CP's changes as 1\,$\rightarrow$\, 3\,$\rightarrow$\,5\,$\rightarrow$\,4\,$\rightarrow$\,2
\,$\rightarrow$\, 0 with increasing $G_V$.

The anomaly terms in Eq.(\ref{eqn:Lagrangian}) are supposed to
originate from the instantons, which are to be screened at
finite chemical potential and temperature\cite{ref:SS}.
Accordingly, both the coupling constants $K$ and $K'$ are
expected to diminish around the phase boundary.
However, we emphasize that the main effect of $K'$ is
to enhance the chiral condensate of the strange quark
and the u-d diquark condensate by each other through the
cubic coupling among them for the realistic quark masses,
and  Figs.\ref{fig:pdGVfixed} and \ref{fig:pdK'fixed} tell us
that even smaller values of $K'$ expected at low temperature and
moderate density can still lead to a quite different phase structure
with multiple CP's when the vector interaction is present under
the charge-neutrality constraint.

\section{The Influence On The Chromomagnetic Instability }
\label{sec: instability }

In this section, we investigate effect of the new axial-anomaly
term  on the chromomagnetic instability of the
asymmetric homogeneous 2CSC phase by varying $K'$.
We shall show that the anomaly-induced
interplay between the chiral and diquark condensates acts toward suppressing
the unstable region of the homogeneous 2CSC phase in the $T$-$\mu$ plane.
Thus the 2CSC phase can become even free from the chromomagnetic instability
provided that $K'$ is larger than a critical value $K'_c$, which
can be reduced significantly when the vector interaction is
incorporated.

The magnetic instability region in the $T$-$\mu$ plane is determined
by calculating the Meissner masses squared which can be negative when the
charge-neutrality constraint is imposed. Here we adopt the same method
as that in \cite{Kiriyama:2006jp} to evaluate the Meissner mass squared
\begin{equation}
m_M^2=\frac{\partial^2}{\partial{B^2}}[\Omega(\Delta)-\Omega(\Delta=0)
]_{B=0}, \label{eqn:Meissner}
\end{equation}
where $B$ has the same meaning as that in \cite{Kiriyama:2006jp}.
Since the strange quark does not take part in the diquark pairing
in the present case, we can directly use the formula for two-flavor
NJL model to calculate the Meissner mass squared.

\begin{figure}
\hspace{-.0\textwidth}
\begin{minipage}[t]{.5\textwidth}
\includegraphics*[width=\textwidth]{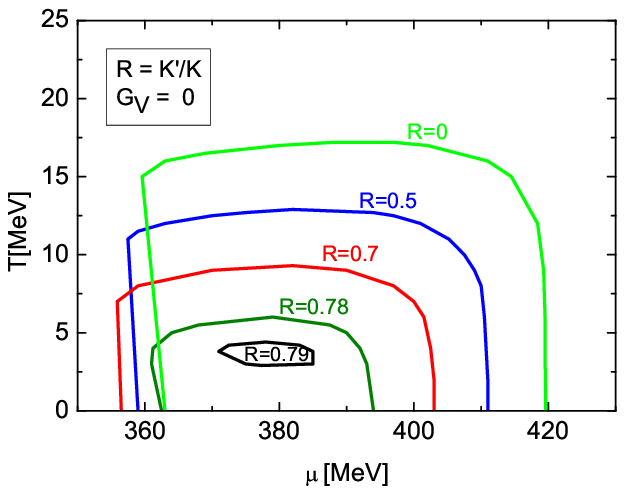}
\centerline{(a)}
\end{minipage}
\hspace{-.05\textwidth}
\begin{minipage}[t]{.5\textwidth}
\includegraphics*[width=\textwidth]{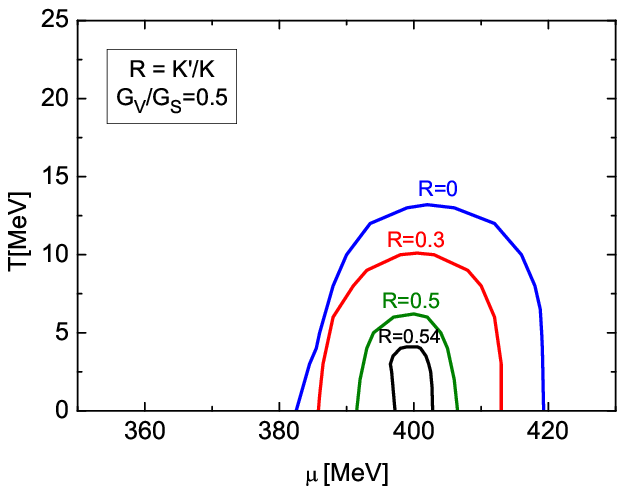}
\centerline{(b)}
\end{minipage}
\caption{
The boundary between the stable and unstable homogenous 2CSC
regions with (right figure) and without (left figure) the vector
interaction in two-plus-one-flavor NJL model. With the increase of
the ratio $K'/K\equiv R$, the unstable region with the chromomagnetic
instability in the $T$-$\mu$ plane shrinks and eventually vanishes.
}
\label{fig:unstable-stable}
\end{figure}

The effect of the coupling constant $K'$ on the chromomagnetic
instability is shown in Fig.\ref{fig:unstable-stable}. We have adopted
the model parameters in Table~\ref{tab:tab1} to calculate the Meissner mass
squared. Figure~\ref{fig:unstable-stable}a displays the change of
the unstable region of the chromomagnetic instability
with varying $K'$ when the vector interaction is not included.
One can see that the instability region tends to shrink with increasing $K'$ and
eventually vanishes for $K'/K>0.8$. This suggests that the neutral homogenous
2CSC phase will be totally free from the chromomagnetic instability
if $K'=K$ that is derived by the Fierz transformation from the usual
instanton vertex. When taking the vector interaction with
$G_V/G_S=0.5$, the unstable region shrinks more significantly with increasing $K'$
 and eventually disappears in the $T$-$\mu$ plane for $K'/K>0.55$,
as shown in Fig.\ref{fig:unstable-stable}b. This could be an expected result
because of the effect of the vector interaction on the instability problem
found in \cite{Zhang:2009mk}.

The reason for the suppression of the chromomagnetic instability by $K'$ is
understood as follows. First of all,  Eq.~(\ref{eqn:mmass}) tells us that the
u-d diquark coupling is enhanced by the presence of the s quark chiral condensate
due to the coupling between the chiral and diquark condensates
induced by the $K'$ term. On the other hand, as was first shown in \cite{Kitazawa:2006zp}
through changing the diquark coupling by hand, the chromomagnetic
instability tends to be suppressed in the strong coupling region and
can be completely gotten rid of when the diquark coupling is strong enough.
Thus we see that the coupling between the u-d diquark and the chiral s-quark
condensates by the  $K'$ term leads to the suppression of the chromomagnetic
instability. This is a new mechanism of the stabilization of the gapless 2CSC
phase, found in the present work. We here emphasize the important role
of the strange quark and the anomaly term in suppressing the instability:
In contrast to the pure two-flavor case, the rather large chiral condensate of
the strange quark enhances the diquark coupling between the u and d quarks
owing to the axial anomaly in the two-plus-one flavor case,
and this enhancement of the diquark coupling causes
the stabilization.

Due to their common effects on the chromomagnetic (in)stability,
Fig.\ref{fig:unstable-stable} suggests that the instability may
be totally cured in the asymmetric homogeneous 2CSC phase when the
coupling constants of the vector interaction and the extended
six-quark interaction are in an appropriate range.
Admittedly, the present work has only dealt with the case of the
so called intermediate diquark coupling. Nevertheless, for a  weaker
 diquark coupling, it is expected that the system can be still free
 from the chromomagnetic instability only with larger couplings for
 both the vector interaction and the anomaly $K'$-term.

%%%%%%%%%%   CONCLUSIONS AND OUTLOOK   %%%%%%%%%%

\section{CONCLUSIONS AND OUTLOOK}

We have explored the phase structure and the chromomagnetic
instability of the strongly interacting matter under the
charge-neutrality constraint within a two-plus-one-flavor
NJL model by incorporating a new anomaly term
as well as the conventional KMT interaction.
The anomaly terms have the forms of six-quark interactions
and violate the $U_A(1)$ symmetry as a reflection of
the axial anomaly of QCD. Similarly to the KMT term, the new
anomaly interaction with the coupling constant $K'$ also induces
a flavor-mixing which leads to a direct coupling between the chiral
and diquark condensates.

We first investigated the role of the axial anomaly on the
emergence of the low or intermediate-temperature CP('s) without the vector
interaction. Owing to the large strange quark mass, the
favored CSC phase near the chiral boundary is 2CSC rather
than CFL, where the electric chemical potential $\mu_e$ required
by the charge-neutrality plays an important role on the chiral phase
transition \cite{Zhang:2008wx}.
The once-declared low-temperature CP
in the symmetric three-flavor limit \cite{Abuki:2010jq} was ruled out in
Ref. \cite{Basler:2010xy} due to the actual dominance of the 2CSC over the
CFL. We have shown that this is true under charge-neutrality constraint without
the vector interaction; the chiral transition in
the low-$T$ region extending zero temperature keeps first order provided that
 $K'$ does not exceed a critical value at which the first-order line completely
disappears.  However, the new chiral anomaly term
enhances the competition between the chiral and diquark condensates under
charge-neutrality constraint, and  gives rise to the intermediate-temperature CP's 
for an appropriate range of $K'$. No such intermediate-temperature CP's had been found 
in the same model when only either the charge-neutrality constraint or the axial 
anomaly is exclusively included, as shown in \cite{Ruester:2005jc} and \cite{Basler:2010xy};
both of which did not take into account the vector interaction, either.

We then investigated the $T$-$\mu$ phase diagram by incorporating the
repulsive vector interaction as well:
We remark that this task may be viewed as an extension of the
work \cite{Zhang:2009mk}, in which the effect of the vector interaction
on the phase diagram is fully explored under the charge-neutrality constraint,
to incorporate the anomaly term. We have found that the cubic coupling between
the chiral and diquark condensates induced by the axial anomaly does not affect
the qualitative results obtained in \cite{Zhang:2009mk}. Rather, the vector
interaction and the anomaly term jointly act so that the
multiple CP's are realized. Indeed, by varying $K'$ with fixed vector coupling
or vise verse, we have shown that all the types of multiple-CP structures
obtained in \cite{Zhang:2009mk} can be reproduced.
In particular, the phase transition in the low-$T$ region extending zero temperature 
becomes a crossover only when the vector interaction is present with a strength 
larger than a critical value. Furthermore, the maximum number
of the CP's can reach as large as five when both the interactions are put on.
In this case, the low- and intermediate-temperature CP's
 can appear even with small values of $K'$
owing to the help by the vector interaction. This is very welcome because
$K'$ in the realistic situation at moderate and high density should be weaker
than that in the vacuum, since the anomaly term is supposed to originate from
the instanton configuration which is expected to be suppressed
at finite density.

Besides the influence on the chiral phase transition, we have shown
that the axial anomaly also plays an important role on the suppression of
the chromomagnetic instability for the asymmetric homogenous 2CSC phase, which
is first disclosed in the present work: With an increase of the extended
six-quark interaction, the $T$-$\mu$ region with the chromomagnetic instability
shrinks and eventually vanishes when the coupling $K'$ is sufficiently large.
In particular, when taking into account the vector interaction simultaneously,
the chromomagnetic instability is suppressed more significantly and can be
completely  gotten rid of by the axial anomaly.

A general and remarkable message obtained from the present investigation is
that the strange quark can significantly affect the properties of the neutral
strongly interacting matter in which the 2CSC phase with u-d pairing is
realized:
Even though the strange quark does not directly participate in the Cooper
pairing in the 2CSC, the interplay between the u-d diquark condensate and
the strange chiral condensate induced by the anomaly term can lead to
drastically different phase structure in the $T$-$\mu$ plane
under charge-neutrality constraint.

It should be remarked here that the contribution of other
possible cubic flavor-mixing terms composed of different
condensates, such as
\begin{equation}
\sigma\rho^{2}=\epsilon^{ijk}\sigma_{i}\rho_{j}\rho_{k},
\end{equation}
which arise from another type of six-quark interaction
\begin{equation}
\mathcal{L}_{\chi{\rho}}^{(6)}\sim \epsilon^{ijk}\epsilon^{lmn}
(\bar{\psi}_i\gamma^{\mu}(1 \pm \gamma_5)\psi_l)(\bar{\psi}_j\gamma_{\mu}(1 \pm \gamma_5)\psi_m)
(\bar{\psi}_k(1 \pm \gamma_5)\psi_n),
\label{eqn:Lagrangian5}
\end{equation}
are all neglected in Eq.~(\ref{eqn:Omega2}) for simplicity.
The interaction~(\ref{eqn:Lagrangian5}) can be derived from the
KMT interaction, which may or may not affect the phase structure.
Beside their direct contribution to the thermodynamic potential, these
flavor-mixing terms also modify the dispersion relations of the
quasi-quarks: For example, the dynamical quark mass becomes dependent on
the quark-number density through the term $\sigma\rho^2$.
It is certainly an interesting problem to explore the
possible effects of these cubic coupling terms on the phase diagram,
we leave such a task to a future work.

Even apart from the neglect of the above vertex (\ref{eqn:Lagrangian5}),
there are some caveats with the present study based on a chiral model
that does not embody the confinement effect, and is relied on the
mean-field approximation. The results obtained in the current study
are largely parameter dependent and bears the shortcomings inherent in
the mean-field approximation. For instance, the result that there can be
multiple CP's associated with the chiral transition and the CSC
actually may merely mean that the QCD matter is very soft for
a simultaneous formation of the diquark and chiral condensates coupled
with the baryonic density along the phase boundary. Of course, a study
which incorporates these fluctuations should be performed, say, by means
of the nonperturbative/functional renormalization group method\cite{Aoki:2000wm,Berges:2000ew}
 with the present model used
as a bare model. More profoundly, the effect of the confinement should be
incorporated even in an effective model approach, which is a more challenging
problem since the mechanism of confinement is still unclear.
Anyway, further studies based on different models and/or methods
are needed to determine whether the low-temperature CP('s) exists.
One of the tasks of future is exploring whether the low-temperature CP('s)
persists or not when the inhomogeneous phases are taken into
consideration such as the chiral crystalline phase
\cite{Nakano:2004cd,Nickel:2009ke,Nickel:2009wj,Carignano:2010ac}
or the LOFF phase \cite{Giannakis:2004pf}.

\acknowledgments
Z.Z. was supported by the Fundamental Research Funds for the Central
Universities of China. T.K. was partially supported by a
Grant-in-Aid for Scientific Research by the Ministry of Education,
Culture, Sports, Science and Technology (MEXT) of Japan (No.20540265),
 by Yukawa International Program for Quark-Hadron Sciences, and by the
Grant-in-Aid for the global COE program `` The Next Generation of
Physics, Spun from Universality and Emergence '' from MEXT.


\begin{thebibliography}{99}

%\cite{Stephanov:2007fk}

%\cite{Cheng:2009be}
\bibitem{Cheng:2009be}
 M.~Cheng {\it et al.},
 %``The finite temperature QCD using 2+1 flavors of domain wall fermions at N_t
 %= 8,''
 Phys.\ Rev.\  D {\bf 81}, 054510 (2010)
 [arXiv:0911.3450 [hep-lat]].
 %%CITATION = PHRVA,D81,054510;%%

%\cite{Borsanyi:2010bp}
\bibitem{Borsanyi:2010bp}
  S.~Borsanyi, Z.~Fodor, C.~Hoelbling, S.~D.~Katz, S.~Krieg, C.~Ratti
  and K.~K.~Szabo,
  %``Is there still any T_c mystery in lattice QCD? Results with physical masses
  %in the continuum limit III,''
  JHEP {\bf 1009}, 073 (2010)
  [arXiv:1005.3508 [hep-lat]].
  %%CITATION = JHEPA,1009,073;%%
%\cite{Endrodi:2011gv}
%\bibitem{Endrodi:2011gv}
  G.~Endrodi, Z.~Fodor, S.~D.~Katz and K.~K.~Szabo,
  %``The QCD phase diagram at nonzero quark density,''
  JHEP {\bf 1104}, 001 (2011)
  [arXiv:1102.1356 [hep-lat]].
  %%CITATION = JHEPA,1104,001;%%


\bibitem{Alford:1998mk}
  M.~G.~Alford, K.~Rajagopal and F.~Wilczek,
  %``Color-flavor locking and chiral symmetry breaking in high density {QCD},''
  Nucl.\ Phys.\  B {\bf 537}, 443 (1999)
  [arXiv:hep-ph/9804403].


\bibitem{Son:1998uk}
 D.~T.~Son,
 %``Superconductivity by long-range color magnetic interaction in  high-density
 %quark matter,''
 Phys.\ Rev.\  D {\bf 59}, 094019 (1999)
 [arXiv:hep-ph/9812287].
 %%CITATION = PHRVA,D59,094019;%%

\bibitem{Schafer:1999jg}
 T.~Sch\"afer and F.~Wilczek,
 %``Superconductivity from perturbative one-gluon exchange in high density
 %quark matter,''
 Phys.\ Rev.\  D {\bf 60}, 114033 (1999)
 [arXiv:hep-ph/9906512].
 %%CITATION = PHRVA,D60,114033;%%

\bibitem{Shovkovy:1999mr}
 I.~A.~Shovkovy and L.~C.~R.~Wijewardhana,
 %``On gap equations and color flavor locking in cold dense QCD with three
 %massless flavors,''
 Phys.\ Lett.\  B {\bf 470}, 189 (1999)
 [arXiv:hep-ph/9910225].
 %%CITATION = PHLTA,B470,189;%%

\bibitem{Schafer:1999fe}
 T.~Sch\"afer,
 %``Patterns of symmetry breaking in QCD at high baryon density,''
 Nucl.\ Phys.\  B {\bf 575}, 269 (2000)
 [arXiv:hep-ph/9909574].
 %%CITATION = NUPHA,B575,269;%%

%
\bibitem{Nambu:1961tp}
  Y.~Nambu and G.~Jona-Lasinio,
  %``Dynamical model of elementary particles based on an analogy with
  %superconductivity. I,''
  Phys.\ Rev.\  {\bf 122}, 345 (1961);\,
Phys.\ Rev.\  {\bf 124}, 246 (1961).

%\cite{Vogl:1991qt}
\bibitem{Vogl:1991qt}
  U.~Vogl and W.~Weise,
  %``The Nambu and Jona Lasinio model: Its implications for hadrons and
  %nuclei,''
  Prog.\ Part.\ Nucl.\ Phys.\  {\bf 27}, 195 (1991).

\bibitem{Klevansky:1992}
%\bibitem{Klevansky:1992qe}
  S.~P.~Klevansky,
  %``The Nambu-Jona-Lasinio model of quantum chromodynamics,''
  Rev.\ Mod.\ Phys.\  {\bf 64}, 649 (1992).

\bibitem{Hatsuda:1994pi}
  T.~Hatsuda and T.~Kunihiro,
  %``QCD phenomenology based on a chiral effective Lagrangian,''
  Phys.\ Rept.\  {\bf 247}, 221 (1994)
  [arXiv:hep-ph/9401310].

\bibitem{Rajagopal:2000wf}
 K.~Rajagopal and F.~Wilczek,
 %``The condensed matter physics of QCD,''
 arXiv:hep-ph/0011333.
 %%CITATION = HEP-PH/0011333;%%

\bibitem{Rischke:2003mt}
 D.~H.~Rischke,
 %``The quark-gluon plasma in equilibrium,''
 Prog.\ Part.\ Nucl.\ Phys.\  {\bf 52}, 197 (2004)
 [arXiv:nucl-th/0305030].
 %%CITATION = PPNPD,52,197;%%

\bibitem{Buballa:2003qv}
 M.~Buballa,
 %``NJL model analysis of quark matter at large density,''
 Phys.\ Rept.\  {\bf 407}, 205 (2005)
 [arXiv:hep-ph/0402234].
 %%CITATION = PRPLC,407,205;%%


\bibitem{Alford:2007xm}
 M.~G.~Alford, A.~Schmitt, K.~Rajagopal and T.~Sch\"afer,
 %``Color superconductivity in dense quark matter,''
 Rev.\ Mod.\ Phys.\  {\bf 80}, 1455 (2008)
 [arXiv:0709.4635 [hep-ph]].
 %%CITATION = RMPHA,80,1455;%%


 \bibitem{Asakawa:1989bq}
 M.~Asakawa and K.~Yazaki,
 %``CHIRAL RESTORATION AT FINITE DENSITY AND TEMPERATURE,''
 Nucl.\ Phys.\  A {\bf 504}, 668 (1989).
 %%CITATION = NUPHA,A504,668;%%

\bibitem{Barducci:1989}
 A.~Barducci, R.~Casalbuoni, S.~De Curtis, R.~Gatto and G.~Pettini,
  %``CHIRAL SYMMETRY BREAKING IN QCD AT FINITE TEMPERATURE AND DENSITY,''
  Phys.\ Lett.\  B {\bf 231}, 463 (1989).

%\cite{Kunihiro:1991hp}
\bibitem{Kunihiro:1991hp}
  T.~Kunihiro,
  %``Chiral restoration, flavor symmetry and the axial anomaly at finite
  %temperature in an effective theory,''
  Nucl.\ Phys.\  B {\bf 351}, 593 (1991).

\bibitem{Berges:1998rc}
 J.~Berges and K.~Rajagopal,
 %``Color superconductivity and chiral symmetry restoration at nonzero  baryon
 %density and temperature,''
 Nucl.\ Phys.\  B {\bf 538}, 215 (1999)
 [arXiv:hep-ph/9804233].
 %%CITATION = NUPHA,B538,215;%%

%\cite{Ruester:2005jc}
\bibitem{Ruester:2005jc}
  S.~B.~Ruester, V.~Werth, M.~Buballa, I.~A.~Shovkovy and D.~H.~Rischke,
  %``The phase diagram of neutral quark matter: Self-consistent treatment of
  %quark masses,''
  Phys.\ Rev.\  D {\bf 72}, 034004 (2005)
  [arXiv:hep-ph/0503184].
  %%CITATION = PHRVA,D72,034004;%%

\bibitem{Abuki:2005ms}
  H.~Abuki and T.~Kunihiro,
  %``Extensive study of phase diagram for charge neutral homogeneous quark
  %matter affected by dynamical chiral condensation: Unified picture for
  %thermal unpairing transitions from weak to strong coupling,''
  Nucl.\ Phys.\  A {\bf 768}, 118 (2006)
  [arXiv:hep-ph/0509172].

\bibitem{Stephanov:2007fk}
As a review, see, M.~A.~Stephanov,
  Prog.\ Theor.\ Phys.\ Suppl.\  {\bf 153}, 139 (2004)
  [Int.\ J.\ Mod.\ Phys.\  A {\bf 20}, 4387 (2005)];\,
%``QCD phase diagram: An overview,''
  PoS {\bf LAT2006}, 024 (2006)
  [arXiv:hep-lat/0701002].
  %%CITATION = POSCI,LAT2006,024;%%

\bibitem{Stephanov:1998dy}
 M.~A.~Stephanov, K.~Rajagopal and E.~V.~Shuryak,
 %``Signatures of the tricritical point in {QCD},''
 Phys.\ Rev.\ Lett.\  {\bf 81}, 4816 (1998)
 [arXiv:hep-ph/9806219].
%
%\cite{Minami:2009hn}
\bibitem{Minami:2009hn}
  Y.~Minami and T.~Kunihiro,
  %``Dynamical Density Fluctuations around QCD Critical Point Based on
  %Dissipative Relativistic Fluid Dynamics-possible fate of Mach cone at the
  %critical point-,''
  Prog.\ Theor.\ Phys.\  {\bf 122}, 881 (2010)
  [arXiv:0904.2270 [hep-th]].
 %
 %\cite{Bowman:2008kc}
\bibitem{Bowman:2008kc}
  E.~S.~Bowman and J.~I.~Kapusta,
  %``Critical Points in the Linear Sigma Model with Quarks,''
  Phys.\ Rev.\  C {\bf 79}, 015202 (2009)
  [arXiv:0810.0042 [nucl-th]].
  %%CITATION = PHRVA,C79,015202;%%

%\cite{Ferroni:2010ct}
\bibitem{Ferroni:2010ct}
  L.~Ferroni, V.~Koch and M.~B.~Pinto,
  %``Multiple Critical Points in Effective Quark Models,''
  Phys.\ Rev.\  C {\bf 82}, 055205 (2010)
  [arXiv:1007.4721 [nucl-th]].
  %%CITATION = PHRVA,C82,055205;%%

\bibitem{Kitazawa:2002bc}
 M.~Kitazawa, T.~Koide, T.~Kunihiro and Y.~Nemoto,
 %``Chiral and color superconducting phase transitions with vector interaction
 %in a simple model,''
 Prog.\ Theor.\ Phys.\  {\bf 108}, 929 (2002)
 [arXiv:hep-ph/0207255].
 %%CITATION = PTPKA,108,929;%%
%
\bibitem{Addenda}
 M.~Kitazawa, T.~Koide, T.~Kunihiro and Y.~Nemoto,
 Prog.\ Theor.\ Phys.\  {\bf 110}, 185 (2003)
[arXiv:hep-ph/0307278v1].
%
\bibitem{Zhang:2008wx}
 Z.~Zhang, K.~Fukushima and T.~Kunihiro,
 %``Number of the QCD critical points with neutral color superconductivity,''
 Phys.\ Rev.\  D {\bf 79}, 014004 (2009)
 [arXiv:0808.3371 [hep-ph]].
 %%CITATION = PHRVA,D79,014004;%%

\bibitem{Zhang:2009mk}
 Z.~Zhang and T.~Kunihiro,
 %``Vector interaction, charge neutrality and multiple chiral critical point
 %structures,''
 Phys.\ Rev.\  D {\bf 80}, 014015 (2009)
 [arXiv:0904.1062 [hep-ph]].
 %%CITATION = PHRVA,D80,014015;%%
%

\bibitem{ref:EHS}
N.~Evans, S.~D.~H.~Hsu and M.~Schwetz, Nucl.~Phys.~{\bf B551}, 275
(1999).

\bibitem{ref:SW-reno}
T.~Sch\"afer and F.~Wilczek, Phys.~Lett. {\bf B450}, 325 (1999).
%


\bibitem{Huang:2004bg}
  M.~Huang and I.~A.~Shovkovy,
  %``Screening masses in neutral two-flavor color superconductor,''
  Phys.\ Rev.\  D {\bf 70}, 094030 (2004)
  \textit{ibid.} D {\bf 70}, 094030 (2004)
  [arXiv:hep-ph/0408268].
%

%\bibitem{newvertex}
%\cite{Rapp:1999qa}
\bibitem{Rapp:1999qa}
 R.~Rapp, T.~Schafer, E.~V.~Shuryak and M.~Velkovsky,
  %``High-density QCD and instantons,''
  Annals Phys.\  {\bf 280}, 35 (2000)
  [arXiv:hep-ph/9904353].

%\cite{Steiner:2005jm}
\bibitem{Steiner:2005jm}
 A.~W.~Steiner,
 %``The color-superconducting t'Hooft interaction,''
 Phys.\ Rev.\  D {\bf 72}, 054024 (2005)
 [arXiv:hep-ph/0506238].
 %%CITATION = PHRVA,D72,054024;%%


\bibitem{Hatsuda:2006ps}
 T.~Hatsuda, M.~Tachibana, N.~Yamamoto and G.~Baym,
 %``New critical point induced by the axial anomaly in dense QCD,''
 Phys.\ Rev.\ Lett.\  {\bf 97}, 122001 (2006)
 [arXiv:hep-ph/0605018].
 %%CITATION = PRLTA,97,122001;%%

\bibitem{Yamamoto:2007ah}
 N.~Yamamoto, M.~Tachibana, T.~Hatsuda and G.~Baym,
 %``Phase structure, collective modes, and the axial anomaly in dense QCD,''
 Phys.\ Rev.\  D {\bf 76}, 074001 (2007)
 [arXiv:0704.2654 [hep-ph]].
 %%CITATION = PHRVA,D76,074001;%%

\bibitem{Baym:2008me}
 G.~Baym, T.~Hatsuda, M.~Tachibana and N.~Yamamoto,
 %``The axial anomaly and the phases of dense QCD,''
 J.\ Phys.\ G {\bf 35}, 104021 (2008)
 [arXiv:0806.2706 [nucl-th]].
 %%CITATION = JPHGB,G35,104021;%%

\bibitem{Abuki:2010jq}
  H.~Abuki, G.~Baym, T.~Hatsuda and N.~Yamamoto,
  %``The NJL model of dense three-flavor matter with axial anomaly:
  %the low
  %temperature critical point and BEC-BCS diquark crossover,''
  Phys.\ Rev.\  D {\bf 81}, 125010 (2010)
  [arXiv:1003.0408 [hep-ph]].
  %%CITATION = PHRVA,D81,125010;%%

%\cite{Basler:2010xy}
\bibitem{Basler:2010xy}
  H.~Basler and M.~Buballa,
  %``Role of two-flavor color superconductor pairing in a three-flavor
  %Nambu--Jona-Lasinio model with axial anomaly,''
   Phys.\ Rev.\  D {\bf 82}, 094004 (2010)
   arXiv:1007.5198 [hep-ph].
  %%CITATION = ARXIV:1007.5198;%%

\bibitem{Alford:1997zt}
 M.~G.~Alford, K.~Rajagopal and F.~Wilczek,
 %``QCD at finite baryon density: Nucleon droplets and color
 %superconductivity,''
 Phys.\ Lett.\  B {\bf 422}, 247 (1998)
 [arXiv:hep-ph/9711395].
 %%CITATION = PHLTA,B422,247;%%


\bibitem{Rapp:1997zu}
 R.~Rapp, T.~Sch\"afer, E.~V.~Shuryak and M.~Velkovsky,
 %``Diquark Bose condensates in high density matter and instantons,''
 Phys.\ Rev.\ Lett.\  {\bf 81}, 53 (1998)
 [arXiv:hep-ph/9711396].
 %%CITATION = PRLTA,81,53;%%


\bibitem{Giannakis:2004pf}
  I.~Giannakis and H.~C.~Ren,
  %``Chromomagnetic instability and the LOFF state in a two flavor color
  %superconductor,''
  Phys.\ Lett.\  B {\bf 611}, 137 (2005)
  [arXiv:hep-ph/0412015].

\bibitem{Gorbar:2005rx}
  E.~V.~Gorbar, M.~Hashimoto and V.~A.~Miransky,
  %``Gluonic phase in neutral two-flavor dense QCD,''
  Phys.\ Lett.\  B {\bf 632}, 305 (2006)
  [arXiv:hep-ph/0507303];
%\bibitem{Gorbar:2005tx}
%  E.~V.~Gorbar, M.~Hashimoto and V.~A.~Miransky,
  %``Neutral LOFF state and chromomagnetic instability in two-flavor dense
  %QCD,''
  Phys.\ Rev.\ Lett.\  {\bf 96}, 022005 (2006)
  [arXiv:hep-ph/0509334];
%\bibitem{Gorbar:2007vx}
%  E.~V.~Gorbar, M.~Hashimoto and V.~A.~Miransky,
  %``Gluonic phases, vector condensates, and exotic hadrons in dense QCD,''
  Phys.\ Rev.\  D {\bf 75}, 085012 (2007)
  [arXiv:hep-ph/0701211].

\bibitem{Kiriyama:2006jp}
  O.~Kiriyama,
  %``Chromomagnetic instability in two-flavor quark matter at nonzero
  %temperature,''
  Phys.\ Rev.\  D {\bf 74}, 114011 (2006)
  [arXiv:hep-ph/0609185]

\bibitem{He:2007cn}
L.~He, M.~Jin and P.~Zhuang,
  %``Neutral color superconductivity including inhomogeneous phases at  finite
  %temperature,''
  Phys.\ Rev.\  D {\bf 75}, 036003 (2007)
  [arXiv:hep-ph/0610121].

%\cite{Fukushima:2005cm}
\bibitem{Fukushima:2005cm}
  K.~Fukushima,
  %``Analytical and numerical evaluation of the Debye and Meissner masses in
  %dense neutral three-flavor quark matter,''
  Phys.\ Rev.\  D {\bf 72}, 074002 (2005)
  [arXiv:hep-ph/0506080].
  %%CITATION = PHRVA,D72,074002;%%

\bibitem{Kitazawa:2006zp}
  M.~Kitazawa, D.~H.~Rischke and I.~A.~Shovkovy,
  %``Stable gapless superconductivity at strong coupling,''
  Phys.\ Lett.\  B {\bf 637}, 367 (2006)
  [arXiv:hep-ph/0602065].
  %%CITATION = PHLTA,B637,367;%%

\bibitem{Klimt:1989pm}
 S.~Klimt, M.~Lutz, U.~Vogl and W.~Weise,
 %``GENERALIZED SU(3) NAMBU-JONA-LASINIO MODEL. Part. 1. MESONIC MODES,''
 Nucl.\ Phys.\  A {\bf 516}, 429 (1990).

%\cite{Kunihiro:1991qu}
\bibitem{Kunihiro:1991qu}
  T.~Kunihiro,
  %``Quark number susceptibility and fluctuations in the vector channel at high
  %temperatures,''
  Phys.\ Lett.\  B {\bf 271}, 395 (1991).

\bibitem{Kobayashi:1970ji}
 M.~Kobayashi and T.~Maskawa,
 %``Chiral symmetry and eta-x mixing,''
 Prog.\ Theor.\ Phys.\  {\bf 44}, 1422 (1970).
 %%CITATION = PTPKA,44,1422;%%

\bibitem{'t Hooft:1976fv}
 G.~'t Hooft,
 %``Computation of the quantum effects due to a four-dimensional
 %pseudoparticle,''
 Phys.\ Rev.\  D {\bf 14}, 3432 (1976)
 [Erratum-ibid.\  D {\bf 18}, 2199 (1978)].
 %%CITATION = PHRVA,D14,3432;%%
 G.~'t Hooft,
 %``How Instantons Solve the U(1) Problem,''
 Phys.\ Rept.\  {\bf 142}, 357 (1986).
 %%CITATION = PRPLC,142,357;%%
%
%\cite{Kunihiro:1989my}
\bibitem{Kunihiro:1989my}
  T.~Kunihiro,
  %``Effects Of The U(A)(1) Anomaly On The Quark Condensates And Meson
  %Properties At Finite Temperature,''
  Phys.\ Lett.\  B {\bf 219}, 363 (1989).
%

%\cite{Fu:2007xc}
\bibitem{Fu:2007xc}
  W.~j.~Fu, Z.~Zhang and Y.~x.~Liu,
  %``2+1 Flavor Polyakov--Nambu--Jona-Lasinio Model at Finite Temperature and
  %Nonzero Chemical Potential,''
  Phys.\ Rev.\  D {\bf 77}, 014006 (2008)
  [arXiv:0711.0154 [hep-ph]].
  %%CITATION = PHRVA,D77,014006;%%

%\cite{Kunihiro:2009ds}
\bibitem{Kunihiro:2009ds}
  T.~Kunihiro,
  %``$X$ Meson aka $\eta'$ and Kobayashi-Maskawa-'t Hooft Six-quark Vertex --
  %$U(1)_A$ Anomaly and Generalized Nambu-Jona-Lasinio Model --,''
  Prog.\ Theor.\ Phys.\  {\bf 122}, 255 (2009)
  [arXiv:0907.3808 [hep-ph]].
%
%\cite{Weinberg:1975ui}
\bibitem{Weinberg:1975ui}
  S.~Weinberg,
  %``The U(1) Problem,''
  Phys.\ Rev.\  D {\bf 11}, 3583 (1975).
%

\bibitem{ref:SS}
T.~Sch\"afer and E.~Shuryak, Rev.~Mod.~Phys. {\bf 70}, 323 (1998).

%\cite{Rehberg:1995kh}
\bibitem{Rehberg:1995kh}
  P.~Rehberg, S.~P.~Klevansky and J.~Hufner,
  %``Hadronization in the SU(3) Nambu-Jona-Lasinio model,''
  Phys.\ Rev.\  C {\bf 53}, 410 (1996)
  [arXiv:hep-ph/9506436].
  %%CITATION = PHRVA,C53,410;%%

\bibitem{Alford:2002kj}
  M.~Alford and K.~Rajagopal,
  %``Absence of two-flavor color superconductivity in compact stars,''
  JHEP {\bf 0206}, 031 (2002)
  [arXiv:hep-ph/0204001];
%\bibitem{Steiner:2002gx}
  A.~W.~Steiner, S.~Reddy and M.~Prakash,
  %``Color-neutral superconducting quark matter,''
  Phys.\ Rev.\  D {\bf 66}, 094007 (2002)
  [arXiv:hep-ph/0205201].

%\cite{Nakano:2004cd}
\bibitem{Nakano:2004cd}
  E.~Nakano and T.~Tatsumi,
  %``Chiral symmetry and density wave in quark matter,''
  Phys.\ Rev.\  D {\bf 71}, 114006 (2005).
  [arXiv:hep-ph/0411350].

\bibitem{Nickel:2009ke}
  D.~Nickel,
  %``How many phases meet at the chiral critical point?,''
  Phys.\ Rev.\ Lett.\  {\bf 103}, 072301 (2009)
  [arXiv:0902.1778 [hep-ph]].
  %%CITATION = PRLTA,103,072301;%%

\bibitem{Nickel:2009wj}
  D.~Nickel,
  %``Inhomogeneous phases in the Nambu-Jona-Lasino and quark-meson model,''
  Phys.\ Rev.\  D {\bf 80}, 074025 (2009)
  [arXiv:0906.5295 [hep-ph]].
  %%CITATION = PHRVA,D80,074025;%%
%
%\cite{Carignano:2010ac}
\bibitem{Carignano:2010ac}
  S.~Carignano, D.~Nickel and M.~Buballa,
  %``Influence of vector interaction and Polyakov loop dynamics on inhomogeneous
  %chiral symmetry breaking phases,''
  Phys.\ Rev.\  D {\bf 82}, 054009 (2010).
%  [arXiv:1007.1397 [hep-ph]].

\bibitem{Shovkovy:2004me}
I. Shovkovy and M. Huang,
\newblock Phys. Lett. B {\bf 564}, 205 (2003); M. Huang and I.
Shovkovy, Nucl. Phys. A {\bf 729}, 853 (2003).

\bibitem{ref:RWP}
C. D. Roberts and A. G. Williams, Prog. Part. Nucl. Phys. 33
(1994) 477; P. C. Tandy, Prog. Part. Nucl. Phys. 39 (1997) 117.

%\cite{Aoki:2000wm}
\bibitem{Aoki:2000wm}
  K.~Aoki,
  %``Introduction to the nonperturbative renormalization group and its recent
  %applications,''
  Int.\ J.\ Mod.\ Phys.\  B {\bf 14} (2000) 1249.

%\cite{Berges:2000ew}
\bibitem{Berges:2000ew}
  J.~Berges, N.~Tetradis, C.~Wetterich,
  %``Nonperturbative renormalization flow in quantum field theory and statistical physics,''
  Phys.\ Rept.\  {\bf 363 } (2002)  223-386.



\end{thebibliography}
\end{document}